\documentclass[prb]{revtex4}

\makeatletter
\def\@dotsep{4.5}
\makeatother

\usepackage{graphicx}
\usepackage{dcolumn}
\usepackage{bm}
\usepackage[active]{srcltx}
\usepackage{amsmath}
\usepackage{amssymb}
\usepackage{epsf}
\usepackage{subfigure}
\usepackage{graphics}
\usepackage{rotating}
\usepackage{bm}
\usepackage{psfrag}
\usepackage{xcolor}

\newcommand{\rP}{\mathrm{P}}
\newcommand{\re}{\mathrm{e} }
\newcommand{\cD}{\mathcal{D}}

\newcommand{\rs}{\mathrm{s}}

\newcommand{\eref}[1]{Eq.~\eqref{#1}}
\newcommand{\Eref}[1]{Equation~\eqref{#1}}
\newcommand{\fref}[1]{Fig.~\ref{#1}}
\newcommand{\Fref}[1]{Figure~\ref{#1}}

\newcommand{\sref}[1]{Sec.~\ref{#1}}

\newcommand{\aref}[1]{Appendix~\ref{#1}}

\begin{document}

\title{Waste-recycling Monte Carlo with optimal estimates: application to free energy calculations in alloys.}

\author{Gilles Adjanor}

\affiliation{Groupe M\'etallurgie, MMC, EDF, Les Renardi\`eres, 77818 Moret-sur-Loing, France}

\author{Manuel Ath{\`{e}}nes}
\affiliation{CEA, DEN, Service de Recherches de M\'etallurgie Physique, 91191 Gif-sur-Yvette, France}

\author{Jocelyn M. Rodgers}

\affiliation{Physical Biosciences Division, Lawrence Berkeley National Laboratory Berkeley, CA 94720, USA}

\begin{abstract}

  The estimator proposed recently by Delmas and Jourdain for
  waste-recycling Monte Carlo achieves variance reduction optimally
  with respect to a control variate that is evaluated directly using
  the simulation data.  Here, the performance of this estimator is
  assessed numerically for free energy calculations in generic binary
  alloys and compared to those of other estimators taken from the
  literature.  A systematic investigation with varying simulation
  parameters of a simplified system, the anti-ferromagnetic Ising
  model, is first carried out in the transmutation ensemble using
  path-sampling. We observe numerically that (i) the variance of the
  Delmas-Jourdain estimator is indeed reduced compared to that of
  other estimators; and that (ii) the resulting reduction is close to
  the maximal possible one, despite the inaccuracy in the estimated
  control variate.  More extensive path-sampling simulations involving
  a FeCr alloy system described by a many-body potential additionally
  show that (iii) gradual transmutations accommodate the atomic
  frustrations, thus alleviating the numerical ergodicity issue present in
  numerous alloy systems and eventually enabling the determination of
  phase coexistence conditions.


\end{abstract}


\maketitle

\section{Introduction}
In Monte Carlo simulations of multi-particle systems, thermodynamic
quantities are traditionally estimated by an arithmetic mean taken
over the Markov chain of states constructed by the simulation. As a
result, the information pertaining to the trial states that have been
generated and rejected by the Metropolis-Hasting algorithm is lost. In
contrast, waste-recycling Monte Carlo is a technique which includes
this information in the estimate in order to improve its
accuracy.~\cite{frenkel:2006}

Waste-recycling estimators have been provided for quantum Monte Carlo
schemes,~\cite{ceperley:1977} multi-proposal
algorithms,~\cite{frenkel:2004} parallel tempering (with pair-replica
exchanges~\cite{coluzza:2005} or multiple-replica
exchanges~\cite{athenes:2008c}),
path-sampling~\cite{athenes:2002b,athenes:2004} and path-sampling with
multi-proposal.\cite{athenes:2002a,boulougouris:2005,athenes:2007,athenes:2010}
In most implementations, waste-recycling was shown to be beneficial
for estimating free energies or potentials of mean force, in the sense
that variance reduction was observed numerically. Cases of variance
augmentation have however been
reported,~\cite{delmas:2006,athenes:2007} showing that the additional
recycled information is not always relevant.

Nevertheless, Delmas and Jourdain demonstrated mathematically that
variance reduction is indeed achieved in the asymptotic limit when the
acceptance function is symmetric to the identity of the old and the
proposed state, as is the case for the sampling algorithm named after
Boltzmann, Glauber or Barker.  The estimator including information
from both the old and the proposed state is then the associated
conditional expectation.~\cite{delmas:2009} The authors also cast
waste-recycling Monte Carlo into a general control variate problem
structure and derived the optimal choice of the control variate in
terms of asymptotic variance.

The purpose of this article is to investigate the relevance of the
optimal estimator for calculating free energy differences, a crucial
task of molecular simulation.~\cite{frenkel:2002} The article is
organized as follows. The multi-particle system is presented formally
in \sref{sxn:system} along with a broader overview of the weighted path
ensemble approach employed, while the more specific aspects concerning
the simulated binary alloys are deferred until \sref{sxn:applications}.
Weighted path ensemble averages are derived in \sref{sxn:pathensemble} and
the Monte Carlo algorithm, including the mono-proposal sampling scheme
and various estimators, are described in \sref{sxn:algorithm}.  In
\sref{sxn:applications}, the methodology is applied to the calculation of
chemical potential differences in binary alloys. While the primary
goal of the study is to compare the statistical variance of the
Delmas-Jourdain estimator with that of other estimators, we also
illustrate the possibilities of the methodology by determining
coexistence conditions using the equal-area construction on the
chemical potential surface or the common-tangent construction on the
reconstructed Gibbs free energy surface.

\section{Extended system and dynamics\label{sxn:system}}

In this article we are interested in calculating the free energy  
difference for converting one particle of an alloy from one 
component type to the other, the systems 0 and 1 discussed below.
  In order to achieve this, we will employ extended systems which
  allow us to convert the Hamiltonian of the system of one composition
  to the Hamiltonian of the system of the other composition via a
  switching parameter $\lambda$, in a manner reminiscent of thermodynamic
  integration or nonequilibrium work calculations.  In this section,
  we introduce much of the basic notation and terminology, and in the
  subsequent section, we discuss how the statistical physics of these
  paths generated while switching $\lambda$ may be used to construct an
  alternate work-weighted path ensemble which is particularly
  well-posed for transmutations between differing alloy compositions.

The free energy of a multi-particle system of interest, labeled 1, can
only be computed as a difference with respect to a reference
thermodynamic state labeled 0. Let $H_1(x)$ and $H_0(x)$ denote the
Hamiltonians of the respective systems with $x \equiv (q,p)$ denoting
the particle positions $q$ and the particle momenta $p$ composing
phase space.  We define an extended Hamiltonian $H(\chi)=(1-\lambda)H_0(x)+\lambda
H_1(x)$ for $0 \leq \lambda \leq 1$ with respect to the extended state
$\chi =(\lambda,x)$. 
The switching parameter $\lambda$ is considered as an additional coordinate for the sake of generality, as the extended Hamiltonian can be given more general forms. 
The phase space associated with all extended
states $\chi$ such that $\lambda = \alpha$ is denoted $\omega_\alpha$,
and the union $\omega=\cup_{0 \leq \lambda \leq 1} \omega_\lambda$
defines the extended phase space. The probability densities in
$\omega_\alpha$ and $\omega$ are written respectively as
\begin{eqnarray}
\rho_\alpha(\chi) & = & \delta_{\lambda-\alpha } (\chi) \exp\left[\beta \left(G_\alpha -H_\alpha(x) \right) \right] \\
\rho(\chi)  &= & \exp\left[\beta \left(G -H(\chi) \right) \right]
\end{eqnarray}
where $G_\alpha$ and $G$ are the Gibbs free energy of the particle
system and of the extended system, respectively, and the inverse
temperature is $\beta$. The pressure $P$ and volume $V$ are implicitly
accounted for by the Hamiltonian. The delta measure
$\delta_{\lambda-\alpha}(d \chi) \equiv \delta_{\lambda-\alpha}(\chi)
d \chi$ is such that under Lebesgue integration $\int_\omega
\varphi(\lambda) \delta_{\lambda-\alpha}(d \chi)=\varphi(\alpha)$ for
any test function $\varphi$.~\cite{LRS2010} The Gibbs free energies
$G_\alpha$ and $G$ are related to each other via the identity
\begin{equation}
 G_\alpha = G - \beta^{-1} \ln \int_{\omega_\alpha} \rho(\chi) d\chi . 
\end{equation}
In the following, the switching parameter will continuously evolve between states $0$ and $1$, but the two values of $\alpha$ of interest for calculating thermodynamic values will be $0$ and $1$. 
The thermodynamical expectation of quantity $\phi$ is expressed as
\begin{eqnarray}
\left\langle \rho_\alpha , \phi \right\rangle =  \int \phi(x) \rho_\alpha (\chi) d\chi . \label{eqn:average}
\end{eqnarray}

\Eref{eqn:average} will be estimated using a waste-recycling Monte Carlo technique from Ref.~\onlinecite{athenes:2007} 
with appropriate modifications to reflect the recently developed Delmas-Jourdain optimal estimator. 
To improve ergodicity in the phase space as well as the accuracy of the estimates, the trial moves are
trajectories generated within a path ensemble using Langevin dynamics
at temperature $\beta^{-1}$ and constant pressure $P$, yielded by
coupling the particles to a thermostat and barostat. In addition, an
external force $f^{\rm ext} _\lambda (x,\lambda)$ depending on the
extended state acts upon the additional variable $\lambda$
mechanically. In the most general set-up, $\lambda$ is equipped with a
mass $m_\lambda$ and evolves according to
\begin{equation}
  m_\lambda \ddot{\lambda} =  f^{\rm ext} _\lambda -\partial_\lambda H \left(x,\lambda \right), \label{eqn:evolution}
\end{equation}
assuming that $\lambda$ is not restricted to the [0,1] interval. 
A path $z = \left\lbrace \chi_t \right\rbrace_{0 \leq t \leq \tau}$ of
duration $\tau$ consists of the sequence of the states $\chi_{0 \leq t
  \leq \tau}$ generated by the Langevin dynamics starting from
$\chi_0$. The conditional probability to generate path $z$ knowing the
initial state $\chi_0$ of the system is denoted by $\rP_{\rm
  fwd}(\chi_{0 \leq t \leq \tau}|\chi_0)$ and can be evaluated numerically from the normal
random deviates used in the stochastic dynamics. The reverse of path
$z$ is denoted by $z^\dagger = \left\lbrace \chi^\dagger_{t}
\right\rbrace_{0 \leq t \leq \tau} $ and $\chi^\dagger_t =
(\lambda,q,-p)_{\tau-t}$ where $(\lambda,q,p)_{t}=\chi_t$. 
The conditional probability to generate path $z$ backward knowing the
final state $\chi_\tau $ of the system is the
probability to generate $z^\dagger$ forward from $\chi^\dagger_0$
\begin{equation}
\rP_{\rm rev}(\chi_{0 \leq t \leq \tau}|\chi_\tau) = \rP_{\rm fwd}(\chi^\dagger_{0 \leq t \leq \tau}|\chi^\dagger_0). 
\end{equation}
and is also evaluable from the normal deviates required to generate $z^\dagger$. 
In the following, the labeling of the probabilities with fwd and rev is dropped and we rather write 
\begin{equation}
\rP(z|\chi_{\alpha\tau}) = \rP_{\rm rev}(\chi_{0 \leq t \leq \alpha \tau}|\chi_{\alpha \tau}) \rP_{\rm fwd}(\chi_{ \alpha \tau\leq t \leq \tau}|\chi_{\alpha \tau}) \nonumber
\end{equation}
The two values of $\alpha$ of interest will be 0 and 1. Hence, the nature of the probabilities is implied by the conditional presence of $\chi_0$ or $\chi_\tau$. 

The forward and reverse conditional probabilities above are two crucial quantities in
waste-recycling Monte Carlo as their ratio enters the acceptance rule
used both by the sampler (\sref{sxn:samplers}) and the estimator
(\sref{sxn:estimators}). The ratio is related to the heat $Q(z)$
transferred from the thermostat and barostat into the system along $z$
via the well-known expression~\cite{Crooks:1998,crooks:2000}
\begin{equation}
 \frac{\rP(z|\chi_\tau) }{ \rP(z|\chi_0) } = \exp \left[\beta Q (z) \right] .\label{eqn:crooks}
\end{equation}
In presence of an external force acting upon $\lambda$, the heat may
be expressed as (see Eq.~26 and Eq.~27 in Ref.~\onlinecite{athenes:2010}):
\begin{equation}
 Q(z) = H(\chi_\tau) - H (\chi_0) - \int_{\lambda(0)}^{\lambda(\tau)}  f^{\mathrm{ext}}_\lambda (\chi_t)  \dot{\lambda} (t) dt\label{eqn:heat}.
\end{equation}
The integral of the product of the external force acting upon
$\lambda$ by the displacement $d \lambda = \dot \lambda dt $
corresponds to the mechanical work $W(z)$ done on the extended
system. As a result, the identity in \eref{eqn:crooks} can be cast into the
equivalent form
\begin{equation}
 \frac{\rP(z|\chi_\tau) \exp \left[-\beta H(\chi_\tau) \right]}
      {\rP(z|\chi_0)    \exp \left[-\beta H(\chi_0   ) \right]} = \exp \left[ - \beta W (z) \right] .\label{eqn:bochkov}
\end{equation} 
that will prove convenient for waste-recycling Monte Carlo because it
contains the Hamiltonians of the distribution(s) of interest. Note
that the identities in \eref{eqn:crooks} or~\eref{eqn:bochkov} remain
valid when the evolution equation \eref{eqn:evolution} is coupled to a
thermal bath via an Ornstein-Uhlenbeck process.~\cite{athenes:2010}

An autonomous scheduling of the additional coordinates based on the
general evolution equation \eref{eqn:evolution} enables the dynamics to
explore all regions of interest while equilibrium information is
retrieved using the ratio in \eref{eqn:bochkov} by waste-recycling Monte
Carlo. This approach is particularly relevant when free-energy is to
be reconstructed in multiple dimensions, or when the additional
coordinates are meta-variables.~\cite{hummer:2001,laio:2002}
Meta-variables are associated with, for instance, a restraining harmonic spring acting upon a generalised or reaction coordinate and characterize the position of a pulling device added to the particle system. Although the additional device modifies the Hamiltonian of interest, its contribution to thermodynamic expectations can be canceled by taking the ensemble average directly in the extended ensemble~\cite{athenes:2010}, defined as follows 
\begin{equation}
 \left\langle \rho, \phi \right\rangle =  \int \phi(x) \rho (\chi) d\chi . \label{eqn:average_ext}
\end{equation}
Meta-variables are most commonly employed when constructing the free energy along reaction coordinates via umbrella sampling, which in its usual implementation~\cite{frenkel:2002} entails correcting for the sampling bias rather than resorting to a marginal expectation like~\eref{eqn:average_ext}. 

Here, we are interested in thermodynamic states 0 and 1 and need not consider~\eref{eqn:average_ext}. 
We further simplify \eref{eqn:evolution} by requiring that $\lambda$ varies at
constant velocity from $\lambda(0)=0$ to $\lambda(\tau)=1$, as
advocated by Jarzynski.~\cite{jarzynski:1997,jarzynski:2008} The
external force satisfying the resulting constraint $\ddot{\lambda}=0$
in \eref{eqn:evolution} is such that $f^\mathrm{ext}_\lambda =
\partial_\lambda H$, hence the work done on the extended system is
\begin{equation}
 W(z)=\int_0^\tau \partial_\lambda H(\lambda_t,x_t,) \dot{ \lambda}_t  dt.  
\end{equation}

\section{Work-weighted path ensemble~\label{sxn:pathensemble}}

  In this section, we show how the extended systems of the
  previous section may be employed to construct the work-weighted path
  ensemble used in our simulations and sampled by waste-recycling
  Monte Carlo.  

  In the following, the full trajectory space $\Omega$ encompasses all paths that connect the $\omega_0$ and $\omega_1$ phase subspace as the additional coordinate $\lambda$ varies monotonically from 0 at $t=0$ to 1 at $t=\tau$. 
  Besides, two parameters $\alpha$ and $\theta$
  are employed which formally may adopt the full range of possible values from 0 to 1.  
  The intent of each of the parameters is
  distinct however.  As previously mentioned, $\alpha$ is meant to indicate
  the thermodynamic states that we are interested in probing, and as
  such the meaningful values of $\alpha$ are strictly 0 and 1.  The
  other parameter $\theta$ is a weighting factor supplied in
  constructing the path ensemble, 
  in a sense indicating for each trajectory the contribution of the two possible generating positions
  $\chi_0=(0,x_0)$ and $\chi_\tau=(1,x_\tau)$ to its associated probability. 
  In the results, a full range of $\theta$ values will 
  be explored, but $\theta=0.5$ is often advantageous,~\cite{athenes:2002b} allowing for 
  a good overlap~\cite{adjanor:2005} with the two work histograms that can be constructed when paths are generated either forward from the equilibrium distribution of system~0 or backward from the equilibrium distribution of system~1. 

The generalized path ensemble has a weighted probability density $\rP_\theta$ for the generating
parameter $\theta$ in $\Omega$ given as~\cite{athenes:2004,adjanor:2005}
\begin{equation}
 \rP_\theta (z) = \re^{ g_\theta } \left\{ \rP(z|\chi_0) \exp\left[-\beta H(\chi_0) \right] \right\}^{1-\theta} \left\{ \rP(z|\chi_\tau)\exp \left[-\beta H(\chi_\tau) \right]\right\}^\theta \label{eqn:sampled}
\end{equation}
where the normalizing constant $g_\theta$ ensures that $\rP_\theta$ is
a probability distribution
\begin{equation}
 \int_\Omega \rP_\theta (z) \cD z =1. \label{eqn:normalization}
\end{equation}
In the following discussion, the notation $\chi_{\alpha \tau}$
characterizes the state of path $z$ at $t=\alpha \tau$, allowing us to
develop the very similar equations for the states of interest,
$\chi_0$ and $\chi_\tau$, with more compact notation.  For the two
particular values $\alpha \in \left\{0,1 \right\}$,
\eref{eqn:normalization} is equivalent to
\begin{eqnarray}
\re^{  g_\alpha} \int_{\omega_\alpha} \re^{-\beta H(\chi)} d\chi = 1  \label{eqn:particular}
\end{eqnarray}
which results from the simplification of \eref{eqn:sampled} into 
$ \rP_\alpha (z) = \re^{ g_\alpha } \re^{-\beta H(\chi_{\alpha \tau})} \rP(z|\chi_{\alpha \tau})$ 
and from the normalization of the conditional probabilities 
\begin{equation}
\int_{\Omega(\chi_{ \alpha \tau})}  \rP(z|\chi_{\alpha \tau}) \cD z = 1
\end{equation}
in each subspace $\Omega(\chi_{\alpha \tau})$ consisting of paths going through $\chi_{\alpha \tau}$. 
As a result, the logarithm of \eref{eqn:particular} with $\alpha \in \left\{ 0, 1 \right\}$ reads  
\begin{eqnarray}
 g_\alpha = - \ln \int_{\omega_\alpha} \exp \left[-\beta H(\chi) \right] d\chi = \beta G_\alpha  \label{eqn:integrale},
\end{eqnarray}
indicating that these normalization constants $g_0$ and $g_1$ are
indeed free energies for the two endpoint systems.

The normalization of the conditional probabilities for $\alpha \in
\left\{0,1 \right \}$ also enables one to express the thermodynamic
expectation of any quantity $\phi$ with respect to the path density
$\rP_\alpha$ rather than the state density $\rho_\alpha$ (for
$\alpha=0$ or $1$)
\begin{eqnarray}
 \left\langle \rho_\alpha,\phi \right\rangle & = & \int_{\omega_\alpha} \phi(x) \exp \left[\beta (G_\alpha - H(\chi))\right] d\chi   \nonumber \\
                                             & = &  \int \left( \int_\Omega \delta_\chi({\chi_{\alpha \tau}})\phi(x) \re^{ g_\alpha-\beta H(\chi) } \rP_\alpha (z|\chi_{\alpha \tau}) \cD z \right) d\chi \nonumber \\
                                             & = &  \int \phi(x_{\alpha \tau}) \rP_\alpha (z) \cD z \nonumber \\
                                             & = &  \left\langle \rP_\alpha ,  \phi_{\alpha} \right\rangle \nonumber
\end{eqnarray}
where $z \rightarrow \phi_\alpha (z)$ is the path functional such that  $\phi_\alpha (z) = \phi(x_{\alpha \tau})$. 

Since importance-sampling is achieved with respect to the distribution
$\rP_\theta$ with $0 \leq \theta \leq 1$, we should rather employ to
the formal path-average
\begin{eqnarray}
 \left\langle \rho_\alpha , \phi \right\rangle =   \left\langle \rP_\theta , \phi_{\alpha} \rP_\alpha/\rP_\theta \right\rangle \label{eqn:pathaverage}.
\end{eqnarray}
The probability density ratio in \eref{eqn:pathaverage} can be cast into
the form
\begin{equation}
 \rP_\alpha / \rP_\theta = \re^{\beta(\theta -\alpha)  W +g_\alpha - g_\theta}, \label{eqn:generalizedCrooks}
\end{equation}
obtained after substituting $\alpha$ for $\theta$ in \eref{eqn:sampled},
dividing by \eref{eqn:sampled} while leaving $\theta$ intact, and further
utilizing the connection between the conditional probabilities and $W$
as defined in \eref{eqn:bochkov}.  Furthermore, substituting
$\re^{g_\theta-g_\alpha }$ for $\phi$ and $\phi_\alpha$ in both sides
of \eref{eqn:pathaverage} yields
\begin{eqnarray}
  \re^{g_\theta - g_\alpha } & = & \left\langle \rP_\theta , \re^{\beta (\theta-\alpha) W}  \right\rangle . \label{eqn:main_average}
\end{eqnarray}
The form of \eref{eqn:main_average} enables one to
express \eref{eqn:generalizedCrooks} without the unknown normalizing
constants $g_{\alpha}$ and $g_{\theta}$, as follows
\begin{equation}
  \rP_\alpha / \rP_\theta = \re^{\beta(\theta -\alpha) \beta W }/\left\langle \rP_\theta , \re^{\beta (\theta-\alpha) W}  \right\rangle . \label{eqn:general}
\end{equation}
Inserting the probability ratio in \eref{eqn:general} into the path average of \eref{eqn:pathaverage} finally yields
for $\alpha \in \left\{ 0, 1 \right \}$
\begin{eqnarray}
 \left\langle \rho_\alpha , \phi \right\rangle & = & \left\langle \rP_\theta , \phi_{\alpha} \re^{\beta (\theta-\alpha)W} \right\rangle /
\left\langle \rP_\theta , \re^{\beta (\theta-\alpha)W}  \right\rangle . 
\end{eqnarray}
The ratio involves two computationally tractable forms that can be estimated using Markov chain Monte Carlo methods. 
In particular, the free energy difference $G_1-G_0 = \beta^{-1}(g_1-g_0)$ can be obtained from the relationship
\begin{equation}
 G_1 - G_0 = -\beta^{-1} \ln \frac{\left\langle \rP_\theta, \exp \left[\beta (\theta-1)W \right] \right\rangle }  { \left\langle \rP_\theta , \exp\left[\beta \theta W \right] \right\rangle }  \label{eqn:evaluation}. 
\end{equation}
The effect of the bridging parameter $\theta$ on the numerical
performance is discussed in Refs.~\onlinecite{athenes:2004}
and~\onlinecite{adjanor:2005} and further studied in
\sref{sxn:applications}. Note that other work-biased distributions than
$\rP_\theta$ and other ensemble averages have been proposed in the literature~\cite{sun:2003,oberhofer:2005} and are reviewed in
Ref.~\onlinecite{lechner:2007}. Here, one recovers Jarzynski's
identity by choosing the importance function $\rP_{\theta}$ to be
$\rP_{0}$
\begin{equation}
 G_1 - G_0 = -\beta^{-1} \ln \left\langle \rP_0, \exp \left[-\beta W \right] \right\rangle~\label{eqn:jarzynski}.
\end{equation}
Similarly, Crooks's probability ratio~\cite{crooks:2000} 
\begin{equation}
  \frac{\rP(z|\chi_\tau) \rho_1(\chi_\tau)}{\rP(z|\chi_0)\rho_0(\chi_0)} = \exp \left[\beta \left(G_1-G_0 - W (z) \right) \right]. \label{eqn:work}
\end{equation}
is recovered from \eref{eqn:general} after setting $\theta=0$ and
$\alpha=1$ and then substituting $\mathrm{P} \rho_0$ and $\mathrm{P}
\rho_1$ for $\mathrm{P}_0$ and $\mathrm{P}_1$ respectively.
\Eref{eqn:work} and \Eref{eqn:bochkov} are two mathematically
equivalent identities referring to two distinct physical
frameworks. In \eref{eqn:work}, $W(z)$ is to be interpreted as a
nonequilibrium thermodynamic work, defined as the integral of the
energies $\delta E (\lambda ; x) = H(\lambda+\delta
\lambda,x)-H(\lambda,x)$ transferred to the system as accounted for by
small changes $\delta \lambda$ in the external generalized mechanical
constraint $\lambda$.~\cite{horowitz:2008,jarzynski:2007} Constraining
the system via the control parameter $\lambda$ amounts to specifying
the equilibrium distribution $\rho_\lambda$. 

Within this framework, the identity in \eref{eqn:work}
entails a free energy difference. In contrast, with the physical
prescription of \sref{sxn:system}, the external mechanical force
$f^\mathrm{ext}_\lambda$ acts transiently on $\lambda$ for a duration
$\tau$ and is then switched off. As a result, the system returns to
its equilibrium distribution $\rho(\chi)=\exp \left[G-\beta H(\chi)
\right]$. 
In this alternative framework,~\cite{jarzynski:2007} the reverse
and forward occurrence probabilities of path $z$ in
the ratio in \eref{eqn:bochkov} rather corresponds to a Bochkov-Kuzovlev
identity.~\cite{bochkov77,bochkov81}

Neither of the two above frameworks holds with respect to the
weighted path ensembles with distributions $\rP_\theta$ when $0 <
\theta < 1$ because, in this situation, no Hamiltonian-based equilibrium distribution can be defined 
for the state ensemble. 
The path formalism is merely introduced to facilitate the
exploration of the phase space as will be shown in
\sref{sxn:applications}. We now discuss how to sample $\rP_\theta$ and
to estimate $\langle \rP_\theta, \cdot \rangle$ via a Markov chain
Monte Carlo algorithm.

\section{Waste-recycling Monte Carlo with Markov Webs~\label{sxn:algorithm}}

\subsection{Samplers~\label{sxn:samplers}}

In practice, the distribution $\rP_\theta$ is approximated by a Markov
chain constructed by importance sampling. Any sampler consists of two
steps: (i) starting from a given path $z_{k}$, a trial path
$\tilde{z}_k$ is generated from a probability distribution $q$; (ii)
the trial path is accepted and added to the Markov chain
$z_{k+1}=\tilde{z}_k$ with an adequate probability $p$, otherwise
$z_{k+1}=z_k$.  So as to ensure the convergence of the Markov chain
towards the correct distribution, the transition probability matrix
$P$ must satisfy the following detailed balance equation
\begin{equation}
 P(z_\rs , z) \rP_\theta (z) = P(z , z_\rs) \rP_\theta (z_\rs), \label{eqn:BD}
\end{equation}
where $P(z_\rs,z)$ is the probability to transit from path $z$ to path
$z_\rs$, and vice versa for $P(z,z_\rs)$.

To formalize the sampler, we write the probability to construct the
proposal $\tilde{z}$ from $z$ as $q(\{ z , \tilde{z} \} | z)$ and the
acceptance probability of the proposal as $p(\tilde{z},\{z,\tilde{z}\}
)$, making the set of the proposed path and the original path explicit in
each case. The rejection probability is
$1-p(\tilde{z},\{z,\tilde{z}\})$ and amounts to transiting to $z$. The
transition probability from $z$ to $z_\rs$ is
therefore~\cite{delmas:2009} 
\begin{eqnarray} P(z_\rs,z) =
  \sum_{\tilde{z}} \left[ \delta_{\tilde{z}} (z_\rs) p(\tilde{z},
    \{z,\tilde{z}\}) + \delta_z(z_\rs) \left(1-p(\tilde{z},
      \{z,\tilde{z}\} ) \right) \right] q(\{z,\tilde{z} \} | z),
  \label{eqn:transition} 
\end{eqnarray} 
where $\delta$ is the delta
distribution and we allow for the possibility that $z_\rs$ is either
the old path $z$ or the proposed path $\tilde{z}$. We write $q( \{
z,\tilde{z} \} | \tilde{z})$ to express the backward probability
enabling one to construct the current path $z$ from the proposal
$\tilde{z}$. Then, an acceptance probability that can be constructed
from \eref{eqn:BD} and \eref{eqn:transition} is \begin{equation}
 p^{M} (\tilde{z},\{z,\tilde{z}\}) = \min \left( 1 , \frac{ q(\{ z , \tilde{z} \} | \tilde{z} ) \rP_\theta (\tilde{z})} {q(\{\tilde{z},z \} | z) \rP_\theta (z)} \right).
\end{equation}
This rule, proposed by Metropolis, is traditionally used because it
maximizes the probability of acceptance. In particular, it is always
larger than the acceptance probability of the Barker algorithm given
by
\begin{equation}
 p^{B} (\tilde{z}|\{z,\tilde{z}\}) = \frac{q(\{z,\tilde{z}\}|\tilde{z}) \rP_\theta (\tilde{z})} {q( \{z,\tilde{z}\} | \tilde{z}) \rP_\theta (\tilde{z}) + q(\{z,\tilde{z}\} | z) \rP_\theta (z)}. 
\end{equation}
This acceptance probability is symmetric so that $p^{B}
(\tilde{z}|\{z,\tilde{z}\})=p^{B} (\tilde{z}|\{\tilde{z},z \})$, in
contrast to the Metropolis rule $p^M$. Moreover, $p^B$ appears to be a
posterior likelihood associated with a marginal probability that is
actually sampled by the algorithm (see Ref.~\onlinecite{athenes:2007}
and \aref{app:inference}).

We consider that the sampler is such that the generated trial paths
$\tilde{z}$ share a common state $\chi_{\alpha \tau}$ with the current
path $z$ where $\alpha \in \{0,1\}$. The constructed pairs of such
paths are denoted by $\{z,\tilde{z}\}^\alpha$ and are called webs. The
proposal functions are given by the conditional probabilities
\begin{equation}
 q (\{z, \tilde{z} \}^\alpha| z) = r_\alpha \rP (\tilde{z} |\chi_{\alpha \tau}). 
\end{equation}
The factor $r_\alpha$ relates to the probability to choose $\alpha=0$ or $\alpha=1$. 
We similarly define $q(\{z, \tilde{z} \}^\alpha | \tilde{z}) =
r_\alpha \rP (z |\chi_{\alpha \tau})$ for the reverse move, with the couple 
$(r_0,r_1)$ in practice chosen to be alternately $(0,1)$ and $(1,0)$. 
Note that another possible choice consisting of setting $r_\alpha=\frac{1}{2}$ 
both for $\alpha=0$ and $\alpha=1$ appears less efficient numerically.~\cite{athenes:2007}

The Barker acceptance rule is therefore ($\alpha \in
\{ 0,1 \}$)
\begin{equation}
p^B(\tilde{z}| \{z,\tilde{z}\}^\alpha) =    \frac{\exp \left[ \beta (\alpha-\theta) W(\tilde{z} ) \right]} { \exp \left[ \beta (\alpha-\theta) W({z} ) \right] + \exp \left[ {\beta (\alpha-\theta) W(\tilde{z})} \right] }.
\end{equation}
Acceptance ratios for $0 <\alpha <1$ are derived in the
literature~\cite{adjanor:2005,ytreberg:2006} but are not implemented
here.

Once the sampling process has been completed, the quantities of
interest can be estimated from the constructed chain of webs $\left\{
  z_k,\tilde{z}_k \right\}_{1 \leq k \leq n}$.

\subsection{Estimators~\label{sxn:estimators}}

The standard estimator for the expectation $\left\langle \rP_\theta ,
  f \right\rangle$ of a path functional $f$ from a Markov chain of
paths $\left\{ z_k \right\}_{1 \leq k \leq n}$ distributed under
$\rP_\theta$ is defined as
\begin{equation}
 I_{n} (f) = \frac{1}{n} \sum_{k=1}^n  f(z_k),
\end{equation}
which is, in essence, the straightforward summation of states in the chain. In
applications of \sref{sxn:applications}, the path functional $f$ will be
chosen to be $f=\re^{\beta(\theta-\alpha) W }$.

Waste-recycling estimators are based on conditional
expectations~\cite{delmas:2009} and include information from trial
moves using the acceptance
probability.~\cite{ceperley:1977,frenkel:2006} In the path-ensemble,
the waste-recycling estimator reads
\begin{equation}
I^{\rm WR}_{n} (f)  = \frac{1}{n} \sum_{k=1}^n f(z_k) p(z_k, \{z_k,\tilde{z}_k \})+f(\tilde{z}_k) p(\tilde{z}_k , \{z_k,\tilde{z}_k \}) \label{eqn:waste-recycling-estimator}
\end{equation}
where the nature of the webs indicating a shared state at either
$\alpha=0$ or $\alpha=1$ is omitted for brevity and $\{z_k,\tilde{z}_k
\}_{1 \leq k \leq n}$ is the web chain.

Following Delmas and Jourdain,~\cite{delmas:2009} waste-recycling is
reformulated as a control variate problem consisting of searching the
value $b \in \mathbb{R}$ minimizing the asymptotic variances
$\sigma_\theta (f,bf)^2$ of the estimators $J^{b}_{n} (f)$ defined by
\begin{equation}
  J^{b}_{n} (f)  = (1-b)I_{n} (f) +  bI^{\rm WR}_{n} (f). \label{eqn:estimator_jb}
\end{equation}
Note that with the present notation, we have $J^0_{n}=I_{n}$ and
$J^1_{n}=I^{\rm WR}_{n}$ for the standard and waste-recycling estimators
with respective asymptotic variances $\sigma_{\theta}(f,0)^2$ and
$\sigma_{\theta}(f,f)^2$.  Delmas and Jourdain proved that the
function $b \rightarrow \sigma_\theta (f,bf)^2-\sigma_\theta (f,0)^2$
is a convex quadratic form whose coefficients are easily evaluable
when sampling is performed with the Barker algorithm.  To make this
remarkable feature of $\sigma_\theta(f,bf)^2$ clear, we shall write
$\mathbb{E}_\theta \left[ \varphi (z_1,z_0) \right]$ for the
conditional expectation of function $\varphi$ when $z_0$ is
distributed under the invariant measure $\rP_\theta$ and $z_1$ is a
random deviate constructed from $z_0$ using the sampler whose
transition probability is $P(\cdot ,z_0)$ given in \eref{eqn:transition}.

The obtained quadratic form~\cite{delmas:2009}
\begin{equation}
 \sigma_\theta (f,bf)^2-\sigma_\theta (f,0)^2 = \frac{1}{2} \mathbb{E}_\theta \left[ \left( f\left(z_1\right)-f\left(z_0\right) \right)^2 \right] \times b^2 - 2 \left\{ \langle \rP_\theta,f^2 \rangle - \langle \rP_\theta,f \rangle^2 \right\} \times b \label{eqn:quadratic}
\end{equation}
is valid for Barker sampling only. The coefficient of the first order
term in $b$ is the variance of the observable. The coefficient of the
second order term is a conditional variance since
$\mathbb{E}_\theta{\left[ f(z_1)-f(z_0)\right]}=0$ (we have $\langle
\rP_\theta, f \rangle = \mathbb{E}_\theta \left[f(z_0)\right] =
\mathbb{E}_\theta \left[f(z_1)\right]$).  Since both variances are
strictly positive for non-constant $f$ on $\left\{z / P_\theta (z) > 0
\right\}$, the quadratic form \eref{eqn:quadratic} has a minimum
occurring at the strictly positive value
\begin{equation}
b^\star = \frac{\left\langle \rP_\theta , f^2 \right\rangle-\left\langle \rP_\theta , f \right\rangle^2}{ \frac{1}{2} \mathbb{E}_\theta \left[ \left(f(z_1)-f(z_0)\right)^2 \right]}. \label{eqn:optimal}
\end{equation}
It is moreover proven~\cite{delmas:2009} that $b^\star > 1 $ when
$\sigma(f,f)^2 > 0$. To simplify the notation, the dependence of
$b^\star$ on $\theta$ and $f$ is omitted.

This implies that the value $b^\star$ that minimizes the variance can be estimated from a single run in spite of the fact the variance $\sigma_\theta(f,0)^2$ of the standard estimator can only be evaluated from a sample of estimates $I_n(f)$ taken in the limit of large $n$ 
owing to the central limit theorem.~\cite{duflo:1997} The optimal parameter $b^\star $ defined by \eref{eqn:optimal} can be estimated by 
\begin{equation}
{\cal I}_n (b^\star)  = \frac{I_{n} (f^2)-I_{ n} (f)^2} {\frac{1}{2n} \sum_{k=1}^{n} (f(z_{k+1})-f(z_{k}))^2}. \label{eqn:optimalestimate}
\end{equation}
from the Markov chain.  An alternative procedure to estimate the
optimal parameter \eref{eqn:optimal} consists of including information
about trial paths $\tilde{z}_{k}$. In particular, $\mathbb{E}_\theta
[\phi(z_1,z_0) ]$ can be estimated via waste recycling by
$\dfrac{1}{n}\sum_{k=1}^{n} p^B(\tilde{z}_k| \{z_k,\tilde{z}_k\})
\phi(\tilde{z}_k,z_k)+ p({z}_k| \{z_k,\tilde{z}_k\})
\phi(z_k,{z}_k)$. With $\phi(z_1,z_0) = (f(z_1)-f(z_0))^2$, the
optimal parameter $b^\star$ defined by \eref{eqn:optimal} can be
estimated using
\begin{equation} {\cal I}^{\rm WR}_n (b^\star) = \frac{I^{\rm WR}_{n}
    (f^2)-I^{\rm WR}_{n} (f)^2}{ \frac{1}{2n} \sum_{k=1}^n
    p(\tilde{z}_k| \{z_k,\tilde{z}_k\}) (f(\tilde{z}_{k})-f(z_{k}))^2
  } \label{eqn:optimalestimateWR}.
\end{equation}

Note that the waste-recycling estimator $J_n^b$ is not optimal at
$b^\star$ in terms of asymptotic variance when sampling is achieved
with the Metropolis algorithm. The question whether one should resort
to the optimal estimator and the suboptimal Barker sampler or to the
optimal Metropolis sampler and a suboptimal estimator has not been
answered theoretically and will be addressed numerically here.

\section{Applications~\label{sxn:applications}}

We now apply this waste recycling Monte Carlo on Markov webs to our 
system of interest, binary alloys of varying composition. 
Initially, we shall make clear how the general notation of the 
preceding sections is connected to this problem of interest. 
Subsequently, we delve in to the numerical applications. 
The performance of the optimal estimator is first assessed by 
comparing with that of traditional estimators in a generic but 
realistic binary system with $\mathrm{A}$ and $\mathrm{B}$ atoms 
interacting on a rigid lattice in \sref{sxn:ising}.  The possibilities 
of the methodology are then illustrated by performing off-lattice 
simulations of a model FeCr alloy in \sref{sxn:eam}.

\subsection{Binary alloys}

The computational cell contains $N$ atoms. In the reference state at
$t=0$, we have $N=N_\mathrm{A}+N_\mathrm{B}$ or
$N=N_\mathrm{Fe}+N_\mathrm{Cr}$ atoms where $N_\mathrm{X}$ refers to
the atom number of type X. In the target state at $t=\tau$, the cells
still contains $N$ atoms, but an A or Fe atom has been transmuted into
a B or Cr atom, respectively.  Paths thus consist of artificially
switching the potential energy of a selected atom. The switching will
be performed instantaneously in the AB alloy~(\sref{sxn:ising}), or
gradually using constant-pressure Langevin
dynamics~\cite{adjanor:2005} in the FeCr alloy~(\sref{sxn:eam}).

The infinitely-fast transmutations of \sref{sxn:ising} are carried out
without changing the atomic masses. Hence, the ratio of the reverse
conditional probability, $\rP(z|\chi_\tau)=(N_\mathrm{B}+1)^{-1}$, to
the forward one, $\rP(z|\chi_0)=(N-N_\mathrm{B})^{-1}$, relates to the
exponential of the ideal chemical potential difference
\begin{equation}
  \frac{\rP(z|\chi_\tau)}{\rP(z|\chi_0)} = \frac{N_\mathrm{B}+1}{N-N_\mathrm{B}} = \exp \left[ \beta \Delta \mu^\mathrm{id} \right]  \label{eqn:muid}.
\end{equation}
Plugging \eref{eqn:muid} into \eref{eqn:bochkov} leads to an identity
relating the work $W$ carried out on the system for transmuting an A
atom into a B atom to the Hamiltonian variation
\begin{equation}
 W(z) =  H(\chi_\tau)- H(\chi_0) - \Delta \mu^\mathrm{id} 
\end{equation}
where $\chi_\tau$ only differs from $\chi_0$ by the transmuted
atom. When transmuting a B into an A atom, the quantity $-W(z)$ must
be considered instead. Besides, $\Delta \mu^\mathrm{id}$ acts as a heat 
transfered from a reservoir of A and B atoms into the system. 

\begin{figure}
\includegraphics{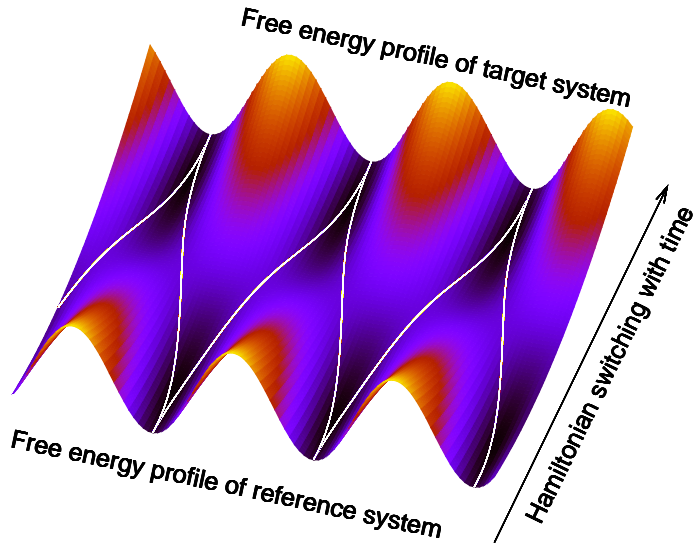}
  \caption{Schematic free energy landscape in the extended space. The
    alloy undergoes forward and reverse atomic transmutations: white
    curves linking free energy minima symbolize the sampled
    transmutations. Path-sampling enables the simulation to explore
    the phase space of both the reference and target systems.}
\label{fig:schematic}
\end{figure}

In contrast, the transmutations of \sref{sxn:eam} are performed
gradually using constant-pressure Langevin
dynamics~\cite{adjanor:2005} and linear Hamiltonian switching. As a
result, the probability ratio \eref{eqn:muid} becomes
\begin{equation}
 \frac{\rP(z|\chi_\tau)}{\rP(z|\chi_0)} = \exp \left[ \beta \Delta \mu^\mathrm{id} + \beta {\cal Q } (z) \right] 
\end{equation}
where ${\cal Q} (z)$ is the heat transferred from the thermostat and barostat
to the particle system,~\cite{adjanor:2005} in addition to the heat $\Delta \mu^\mathrm{id}$ 
transfered from the atomic reservoir.  We thus have
\begin{equation}
 W(z) =  H(\chi_\tau)- H(\chi_0) - \Delta \mu^\mathrm{id} - {\cal Q } (z) 
\end{equation}
In both set-ups, sampling forward and backward transmutations is
sufficient to explore the phase spaces of the alloy. As illustrated in
\fref{fig:schematic}, accepting several paths amounts to exchanging
the allocation of atoms on the underlying lattice of the reference and
target systems.

Simulations aim at extracting differences of chemical potentials
$\Delta \mu$ between both species as a function of the alloy
composition. We have either $\Delta \mu = \mu_\mathrm{B} -
\mu_\mathrm{A}$ or $\Delta \mu = \mu_\mathrm{Cr} - \mu_\mathrm{Fe}$
depending on the involved alloy system.  Both quantities indeed
correspond to the difference of Gibbs free energy between the target
and reference states
\begin{equation}
\Delta \mu = \beta^{-1} (g_1 - g_0) .
\end{equation}
The generic estimator will be  
\begin{equation}
 \mathcal{J}_n^b(\Delta \mu ) = - \beta^{-1} \ln \frac{J_n^b (\re ^{\beta (\theta-1) W})}{J_n^b(\re ^{\beta \theta W})}  \label{eqn:mu_evaluation}
\end{equation}
The numerator and denominator in the logarithm are estimates of $\exp
\left[ g_\theta -g_1 \right]$ and $\exp \left[ g_\theta-g_0 \right]$
obtained simultaneously by sampling the probability measure
$\rP_\theta$ and appling the generalized estimator in \eref{eqn:estimator_jb}.

Let $c=N_\mathrm{B}/N$ or $c=N_\mathrm{Cr}/N$ denote the alloy
concentration in B or Cr. A change in the monotonicity of the function
$c \rightarrow \Delta \mu (c)$ is a signature of phase
coexistence. The conditions of phase equilibria can then be determined
via the equal-area construction with respect to $\Delta \mu(c)$, or
equivalently, via the common-tangent construction with respect to the
Gibbs free energy
\begin{equation}
 G(c) = \int_0^c \Delta \mu (c^\prime) dc^\prime. \label{eqn:gibbs_free_energy}
\end{equation}
measured per atom. 

\subsection{AB system with constant pair interactions~\label{sxn:ising}}

In our simplified binary alloy lattice model, the rigid lattice is
body centered cubic. Interaction energies are taken as pair
interactions $\epsilon_{\mathrm{XY}}$ between nearest-neighbor sites,
where X and Y may equal A or B. The ordering enthalpy
$\epsilon=\epsilon_\mathrm{AA}+\epsilon_\mathrm{BB}-2\epsilon_\mathrm{AB}$
plays a key role as it entirely determines the thermodynamics of the
system.~\cite{ducastelle:1991} Without loss of generality, we can
choose $\epsilon_\mathrm{AA}=\epsilon_\mathrm{AB}=0$ which leads to an
Ising-type internal energy
\begin{equation}
 H(\chi) = \epsilon \sum_{(i,j) \in E} \eta_\mathrm{B}^i \eta_\mathrm{B}^j 
\end{equation}
where the summation runs over the ensemble $E$ of nearest neighbor
pairs only and $\eta_\mathrm{B}^i$ is 1 or 0 depending on whether site
$i$ is occupied by a B atom or not. The cell contains $N=2^{12}$ sites
with $N_\mathrm{B}$ atoms of type B at $t=0$.

We set $\epsilon = -30$~meV. With a negative ordering energy, the
system exhibits a miscibility gap below which the solid solution
decomposes into A-rich and B-rich phases. The unmixing transition is
of first-order except for the A$_{0.5}$B$_{0.5}$ composition where it
is second-order~\cite{ducastelle:1991} and where the critical
temperature is $T_c \approx -\epsilon/(k \times
0.62)$.~\cite{athenes:2000}

We first check the Delmas-Jourdain prediction that the asymptotic
variance $\sigma(f,bf)^2$ of $J^b_n(f)$ is minimal at $b=b^\star$ when
the Barker sampler is used. Since $\lim\limits_{n \to + \infty} n
\mathrm{ var}_n (f,bf) = \sigma(f,bf)^2 $ where $\mathrm{var}_n
(f,bf)$ is the statistical variance of $J_n^b(f)$, we evaluate
$\mathrm{var}_n (f,bf)$ as a function of $b$ for large enough $n$ and
confirm whether its minimum occurs close to estimated values of
$b^\star$. Note that $b \rightarrow \mathrm{var}_n (f,bf)$ is a
quadratic form regardless of the value of $n$.

We set temperature to $348$~K, the simulation parameter $\theta$ to
$1/2$, and $\mathrm{B}$ concentration to $50 at.\%$. The two
quantities $\exp\left[g_{1/2} -g_0 \right]$ and $\exp\left[g_{1/2}
  -g_1 \right]$ are equal to each other for this particular set-up,
and have been evaluated via ensemble average \eref{eqn:main_average} and
the estimator $J_n^b(f)$ given in \eref{eqn:estimator_jb} with
$f=\re^{\pm \frac{1}{2}\beta W}$ and $0 \leq b \leq 20$. The
statistical variances $\mathrm{var}_n(f,bf)$ have been computed from
$m=10^7$ estimates generated using distinct random seeds. Each
estimate consists of $n=2\cdot 10 ^{3}$ transmutations, from
$\mathrm{A}$ into $\mathrm{B}$ and from $\mathrm{B}$ into $\mathrm{A}$
alternatively. For each estimation, we also record the estimated value
of $b^\star$ using estimator $\mathcal{I}_n(b^\star)$ given
in \eref{eqn:optimalestimate} and estimator
$\mathcal{I}^{\mathrm{WR}}_n(b^\star)$ given
in \eref{eqn:optimalestimateWR}, and additionally construct their
histograms: $\langle \mathrm{h}^0(b-b^\star) \rangle$ and $\langle
\mathrm{h}^1(b-b^\star) \rangle$ respectively denote the probabilities
that estimates $\mathcal{I}_n(b\star)$ or
$\mathcal{I}^\mathrm{WR}_n(b^\star)$ are equal to $b$.

\begin{figure}
\includegraphics{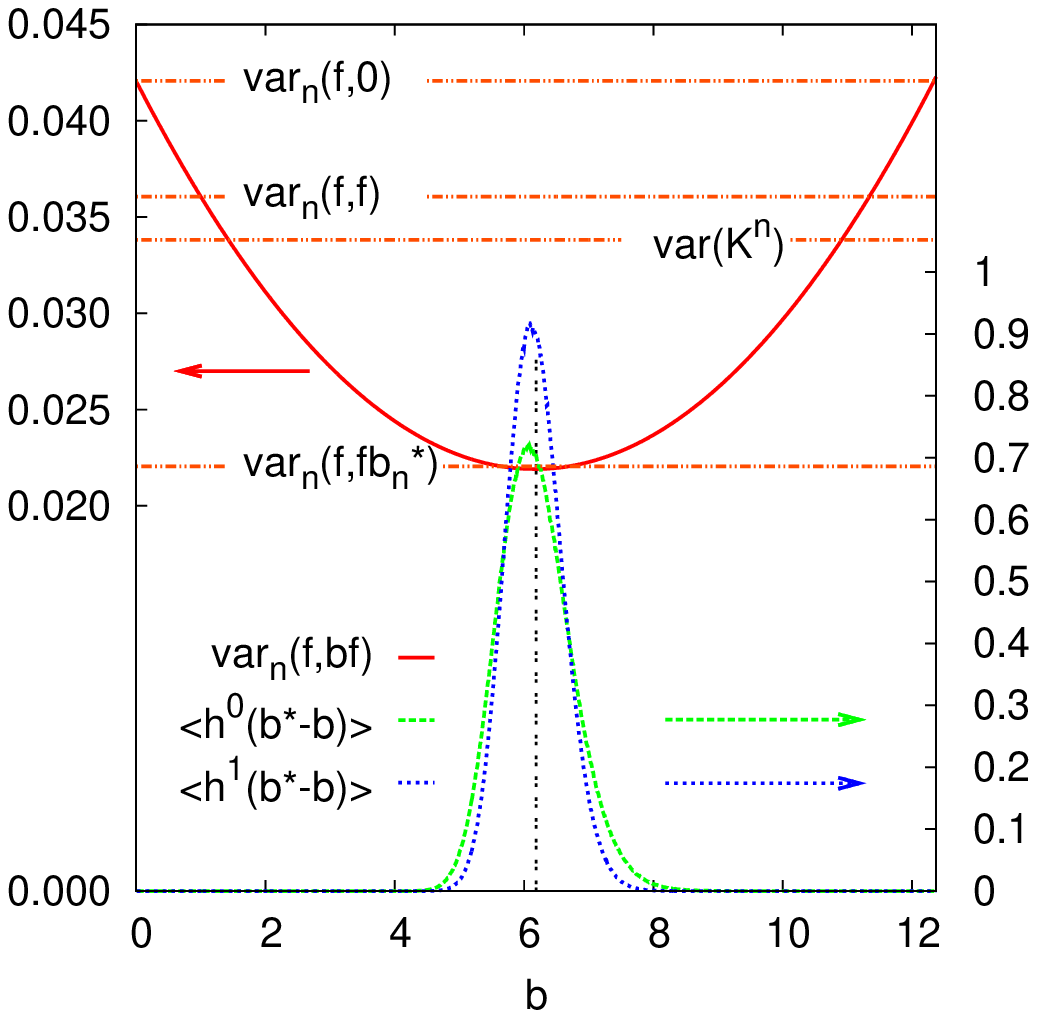}
  \caption{(left ordinate) $\mathrm{var}_n(f,bf)$ is the statistical
    variance of $J_n^b(\re^{-\beta W/2})$ given in
    \eref{eqn:estimator_jb} and is evaluated from $10^7$ estimates.
    Note that $\mathrm{var}_n(f,0)$ and $\mathrm{var}_n(f,f)$
    corresponds to the variances of $I_n(\re^{-\beta W/2})$ and
    $I_n^{\mathrm{WR}}(\re^{-\beta W/2})$.
    $\mathrm{var}_n(f,b_n^\star)$ is defined in the text; (right
    ordinate) $\langle \mathrm{h}^0 (b^\star-b) \rangle$ and $\langle
    \mathrm{h}^1 (b^\star-b) \rangle$ are the histograms of optimal
    paramameter $b^\star$ estimated without or with waste-recycling.}
  \label{fig:quadratic_histo} 
\end{figure}

As shown in \fref{fig:quadratic_histo}, the quadratic form $\mathrm{var}_n
(f,bf)$ is indeed minimum at the value corresponding to the best
$b^\star$-estimate indicated by the vertical double-dotted segment,
obtained by combining the $mn$ available data from either
$\mathcal{I}_{nm}(b^\star)$ or
$\mathcal{I}_{nm}^\mathrm{WR}(b^\star)$. The minimum is also very
close to the horizontal line labeled $\mathrm{var}_n(f,fb^\star_n)$
corresponding to a variance obtained from the $m$ estimates
$J_n^{b^\star_n}(f)$ after substituting the corresponding
$\mathcal{I}_n(b^\star)$ for $b^\star_n$ in each estimate.
Substituting $\mathcal{I}^{\mathrm{WR}}_n(b^\star)$ for $b^\star_n$
further decreases the variance by $0.24\%$, an amount not visible on
the graph. A more noticeable benefit to employing waste-recycling for
estimating $b^\star$ can be seen from the two histograms of $b^\star$
displayed in \fref{fig:quadratic_histo}. Histogram $\langle
\mathrm{h}^1(b-b^\star) \rangle$ is slightly narrower than histogram
$\langle \mathrm{h}^0(b-b^\star) \rangle$ obtained without
waste-recycling.

We also observe in \fref{fig:quadratic_histo} that the optimal
estimator exhibits a smaller statistical variance than that of the
residence-weight estimator $K_n$ (derived in
\aref{app:inference}), although the available theory does not
predict the extent to which the observed tendency is general. We also
don't know the value $\theta^\star$ of the control parameter that
minimizes the total variance in the estimates of $\Delta \mu$
from \eref{eqn:mu_evaluation}. Qualitatively, the symmetric setting
$\theta=1/2$ is often
used~\cite{athenes:2002b,athenes:2004,adjanor:2005,ytreberg:2006}
because the work distribution associated with $\rP_{1/2}$ exhibits
sufficient overlaps with those associated with both $\rP_{0}$ and
$\rP_{1}$, ensuring a fast convergence of the involved exponential
averages.~\cite{adjanor:2005}

\begin{figure}
\includegraphics{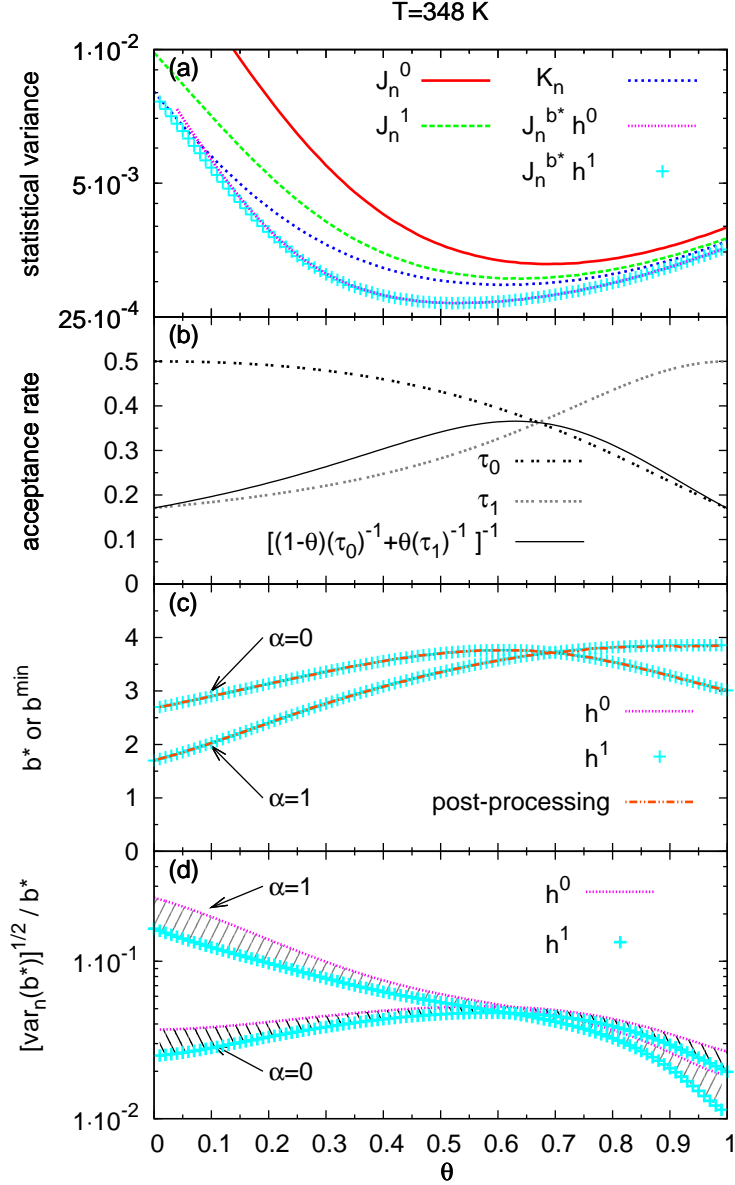}
  \caption{(a) estimator variances; (b) acceptance rates; (c) optimal
    paramater; (d) reduced deviation of $b^\star$ as a function
    $\theta$ [$\mathrm{var}_n(b^\star)$ is the variance of the
    estimates of $b^\star$]. Simulations are performed at $T=348$~K.}
\label{fig:pannel}
\end{figure}

We carried out a series of simulations with varying the value of
$\theta$ in the range [0,1] so as to locate the optimal value
$\theta^\star$ with respect to the evaluation of $\Delta \mu$ using
estimator \eref{eqn:mu_evaluation}, at the asymmetric composition $c=10
at.\%$ B. The statistical variance is obtained using $m=10^7$
independent estimates but with $n=2 \cdot 10^4$ transmutations in each
estimate (10 times more than previously). 
The results in \fref{fig:pannel} demonstrate that the observed
hierarchy in the estimator performance is preserved for any $\theta
\in [0,1]$. The optimal estimators $J_n^{b^\star}$ achieve the best
performance whether waste-recycling is used (h$^1$) or not
($\mathrm{h}^0$) for estimating $b^\star$. Interestingly, using the
optimal estimator shifts the minimum of the variance to a value of
$\theta$ close to $0.5$ from a value about $0.7$ with the standard
estimator. 
Given that $b^\star$ is about 3 when $\theta=1/2$ close to the
minimum of the variance, the ``standard'' contribution involving the Markov chain of paths 
accepted in the sampled distribution $\rP_\theta$ is assigned with a weighting factor $1-b^\star$  close to the negative value of $-2$. The ``waste-recycling'' contribution involves both the accepted and trial paths and is assigned with a weighting factor $b^\star$ that is positive and larger than $1-b^\star$ in absolute value. Hence, compared to the overall contribution of the accepted paths, trial transmutions generated from the forward and backward path distributions contribute to the free-energy estimates quite substantially. 

We have then post-processed the collected data (the $10^7$ estimates)
by evaluating the statistical variances $\mathrm{var}_n(f,bf)$ of
estimator $J_n^b(f)$ (given in \eref{eqn:estimator_jb} with
$f=\re^{\beta(1/2-\alpha) W }$, $\alpha \in \{0,1\}$ and $n=2 \cdot 10
^4$) as a function of $b \in [0 , 20]$. The mimima of the quadratic
forms in $b$ are plotted in panel (c) of \fref{fig:pannel} as a
function of $\theta$ (curve labeled 'post-processing'). As expected
theoretically, they perfectly coincide with the online estimations of
$b^\star$ without waste recycling as in \eref{eqn:optimalestimate} or with
waste recycling as in \eref{eqn:optimalestimateWR}, which are also plotted
for comparison. Panel (d) of \fref{fig:pannel} shows the higher
accuracy obtained for the waste-recycling estimates via
\eref{eqn:optimalestimateWR}.

\begin{figure}
\includegraphics{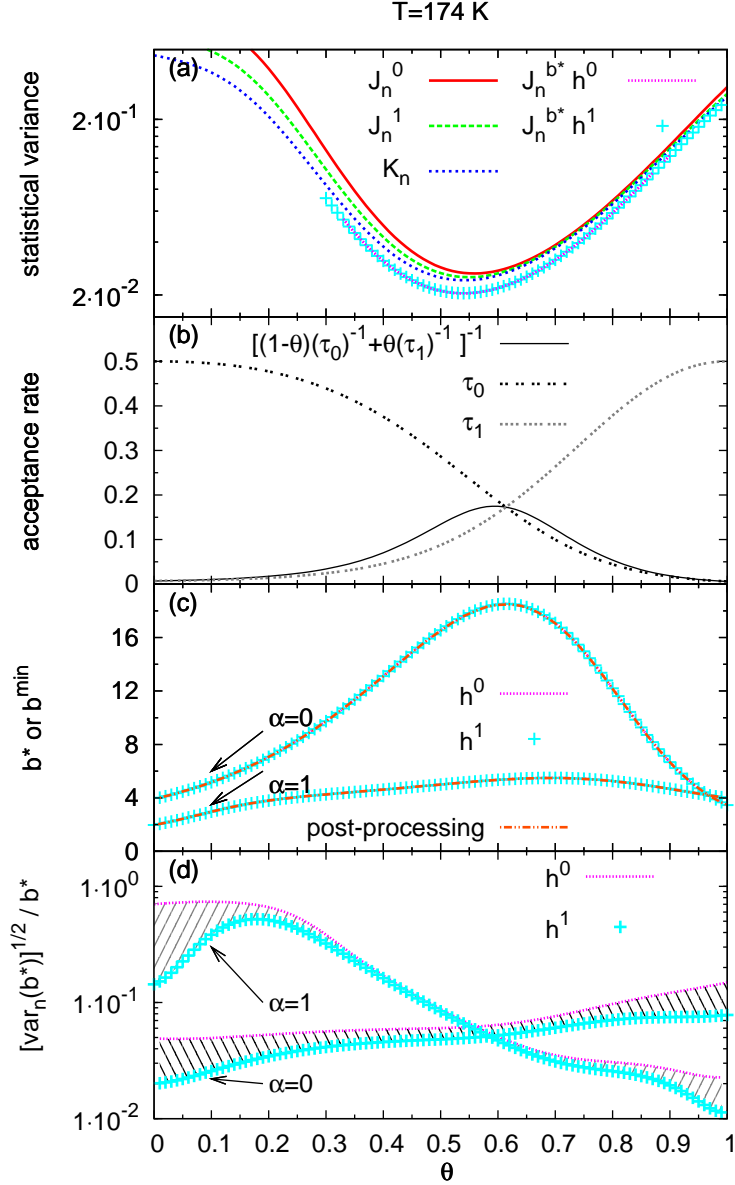}
\caption{Same as in \fref{fig:pannel} but at $T=174$~K.}
\label{fig:pannel_174}
\end{figure}

\begin{figure}
\includegraphics{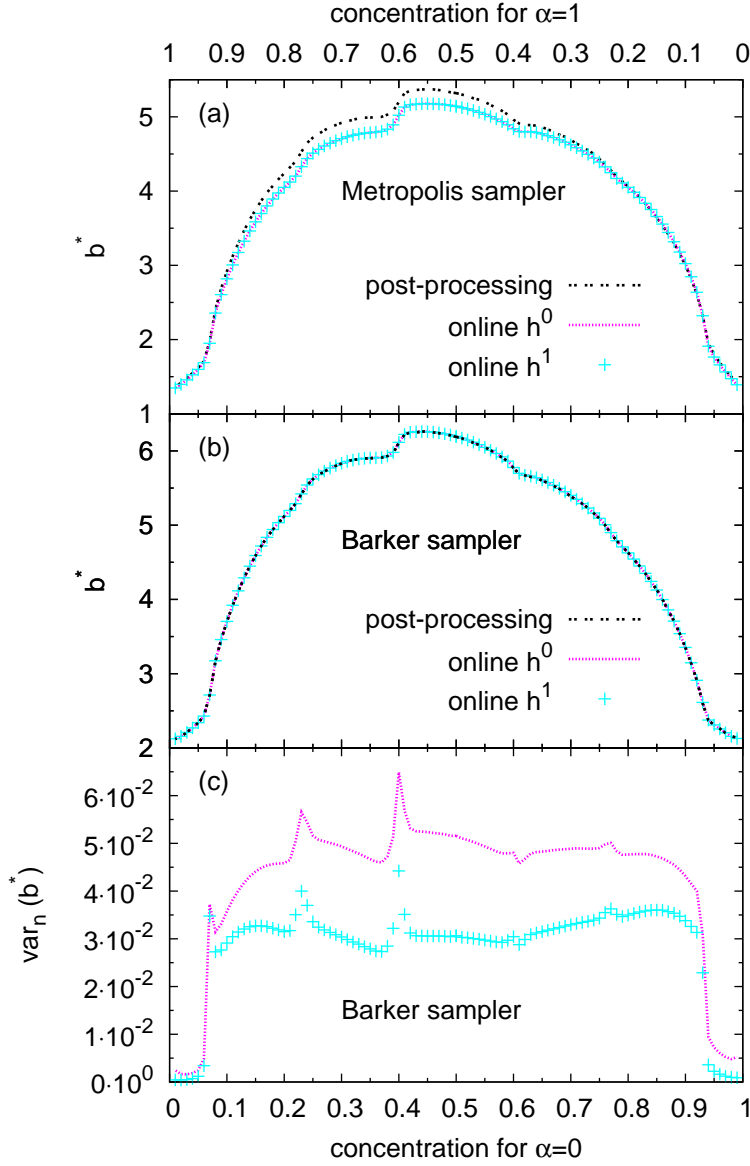}
  \caption{(a) disagreement between $b^\star$ and $b^{\mathrm{min}}$
    obtained from the true minimum after post-processing with the
    Metropolis sampler; (b) agreement between the same quantities with the
    Barker sampler; (c) statistical variances of the $b^\star$
    estimated with the Barker sampler.}
\label{fig:disagreement}
\end{figure}

Additional simulations are carried out at temperature
$\beta^{-1}=174$~K with varying $\theta$ or at $\beta^{-1}=348$~K with
varying concentration (and imposing $\theta=1/2$). Results are
displayed in \fref{fig:pannel_174} and \fref{fig:disagreement},
respectively. The same qualitative trends are observed. Nonetheless,
for the low temperature simulation, much smaller acceptance rates are
measured. A small acceptance rate decreases the denominator in both
\eref{eqn:optimalestimate} and \eref{eqn:optimalestimateWR} explaining
the higher values of $b^\star$ and the difficulties in estimating the
variance when $\theta < 0.25$ where a few negative estimates of
$\re^{g_\theta-g_\alpha}$ prevented evaluation of the logarithm in
\eref{eqn:mu_evaluation}.

We have carried out additional Monte Carlo simulations using the
Metropolis sampler with $\theta=0.5$ and $\beta^{-1}=348$~K and with
varying B concentration in order to compute the minima of the
statistical variances of estimator $J_n^b(\re^{\beta(1/2-\alpha) W })$
as a function of concentration and for $\alpha \in \{0,1\}$.
Similarly, $10^{7}$ estimates have been used. The minima
$b^{\mathrm{min}}$ are plotted as a function of concentration in
\fref{fig:disagreement} (curve labeled 'post-processing') and do not
coincide with the online estimation in \eref{eqn:optimalestimate}
or \eref{eqn:optimalestimateWR}, as expected for the Metropolis sampler.

Because evaluating statistical variances requires a considerable
computational cost, accurately determining $b^{\mathrm{min}}$ by
post-processing from the minimum of the function $b \rightarrow
\mathrm{var}_n(f,bf)$ is more difficult than accurately estimating
$f$, the quantity of interest. Post-processing the data obtained using
the Metropolis sampler is thus not suitable to the simulation of
realistic systems for which accuracy is an issue.

\begin{figure}
\includegraphics{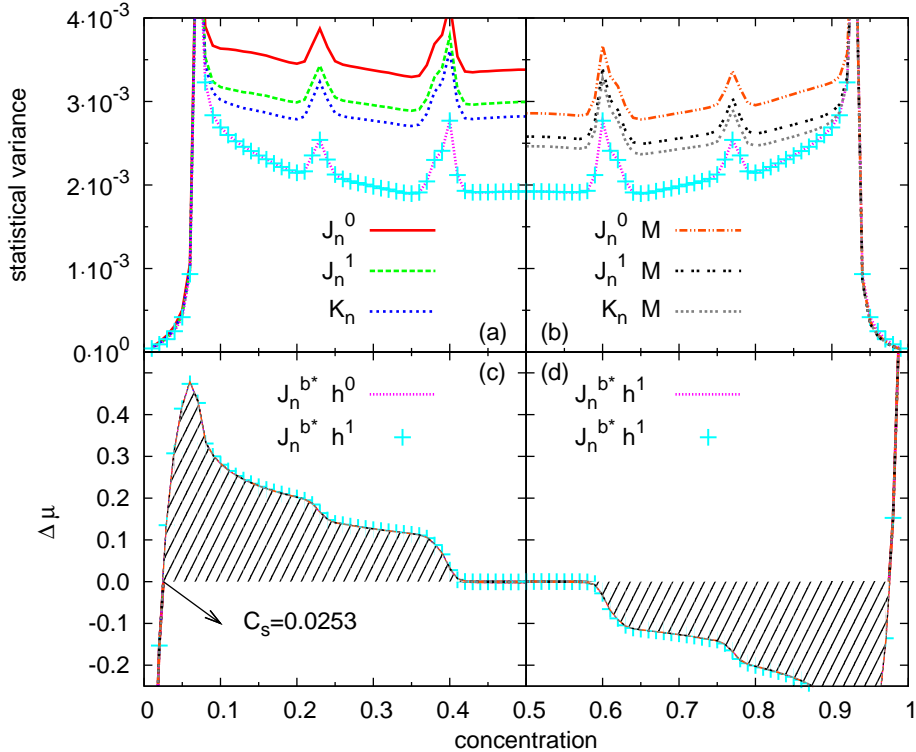}
  \caption{(a,b) statistical variances of estimated differences of
    excess chemical potentials [displayed in (c,d)] as a function of
    concentration. Used estimators are indicated in the legends. The
    symmetry of this Ising system with respect to $C_B=50\%$ explains
    the equality between chemical potentials ($\Delta \mu=0$) at the
    solubility limit $C_s$.}
\label{fig:var_conc}
\end{figure}

The question of whether the combination of the optimal Metropolis
sampler and the suboptimal estimators $I_n$ or $I_n^\mathrm{WR}$ may
achieve better efficiency than the combination of the suboptimal
Barker sampler and the optimal estimator $J_n^{b^\star}$ is not
answered by theory.~\cite{delmas:2009} Therefore, we now address this
numerically. The statistical variances of the $\Delta \mu$-estimates
obtained using the biased estimator of \eref{eqn:mu_evaluation} have been
calculated and plotted as a function of $\mathrm{B}$ concentration in
\fref{fig:var_conc}. We observe that it is always more efficient to
implement the optimized estimator $J_n^{b^\star}$ with the suboptimal
sampler than estimator $J_n^{1}$ with Metropolis sampling, or even the
residence-weight estimator $K_n$ with Metropolis sampling. Note that
$\mathrm{J}_n^{b^\star}$ is still a valid and unbiased estimator when
combined with the Metropolis sampler (but is not optimal anymore).
Here, the combination was found slightly more efficient, with variance
further decreased by 4.7$\%$ for $A_{0.5}B_{0.5}$ compared to Barker
sampling and $\mathrm{J}^{b^\star}_n$. However, variance reduction is
not guaranteed mathematically and cases of variance augmentations have
been reported when implementing waste-recycling estimators with a
Metropolis sampler.~\cite{delmas:2009,athenes:2007}

Hence, the combination of Metropolis samplers with estimation of $b$
is not recommended, is not presented in the plots, and is not further
considered.

\subsection{Fe-Cr system with EAM interactions\label{sxn:eam}}

We now test our estimators on a more difficult model whose
inter-atomic potentials are based on the embedded atom method (EAM).
We have implemented the EAM potentials of Olsson {\it et
  al.}~\cite{olsson:2005} developed to model the $\alpha$ and $\alpha^\prime$ phases of the 
FeCr binary system. While
this EAM potential correctly reproduces the BCC structure of Iron and
Chromium, its ground state (found among 99 candidate BCC structures
based on the theory of the convex polyhedron in the correlation
functions space~\cite{ducastelle:1991}) is an intermetallic structure
with positive ordering enthalpy at $c$=50 at.\% Cr.~\cite{bonny:2009}
So as to avoid the formation of additional phase stability fields, the
intermetallic structure was excluded from the construction of the
complete phase diagram using the Alloy-Theoretic Automated Toolkit
(ATAT) package in Ref.~\onlinecite{bonny:2009}. The ATAT approach involves, 
firstly, constructing effective cluster interaction energies 
on a rigid lattice based on a few relevant structures previously relaxed using the EAM potential 
and, secondly, performing Monte Carlo simulations on the rigid lattice. 
Note however that the expected miscibility gap of FeCr alloy was observed in Monte 
Carlo simulations carried out on a rigid lattice with the 
interaction energies directly deduced from the EAM potential.~\cite{olsson:2005} 
Here, the rigid-lattice assumption is entirely released in the construction of the phase diagram.  

\begin{figure}
\includegraphics[scale=0.5]{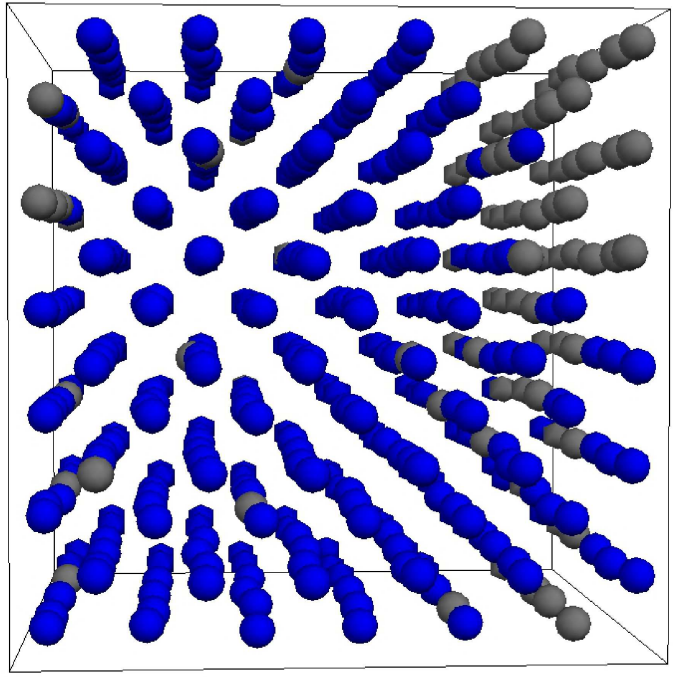}
\caption{Snapshot of a simulated microstructure containing 20\% at. Cr obtained at $T=300$K. Fe and Cr atoms are displayed in blue and gray, respectively.}
\label{fig:snapshot}
\end{figure}

We perform path-sampling simulations to compute the chemical potential
difference with varying concentration and temperature. Transmutations
are now performed gradually in 10$^2$ steps with linear Hamiltonian
switching and constant-pressure Langevin dynamics~\cite{adjanor:2005}
in a computational cell containing 432 atoms. Equilibration proceeds
in 20 transmutations per atom, starting from a random distribution of
the atoms on $6^3$ unit cells of the BCC structure. Simulations have
been carried out with temperature ranging from 300 to 1700~K in step of 25~K. 
Examination of the simulated microstructures show evidende of phase separation at low enough temperatures. 
\Fref{fig:snapshot} displays the snapshot of a typical phase separated microstructure 
for alloy Fe$_{0.8}$Cr$_{0.2}$ at 300~K. 

\begin{figure}
\includegraphics{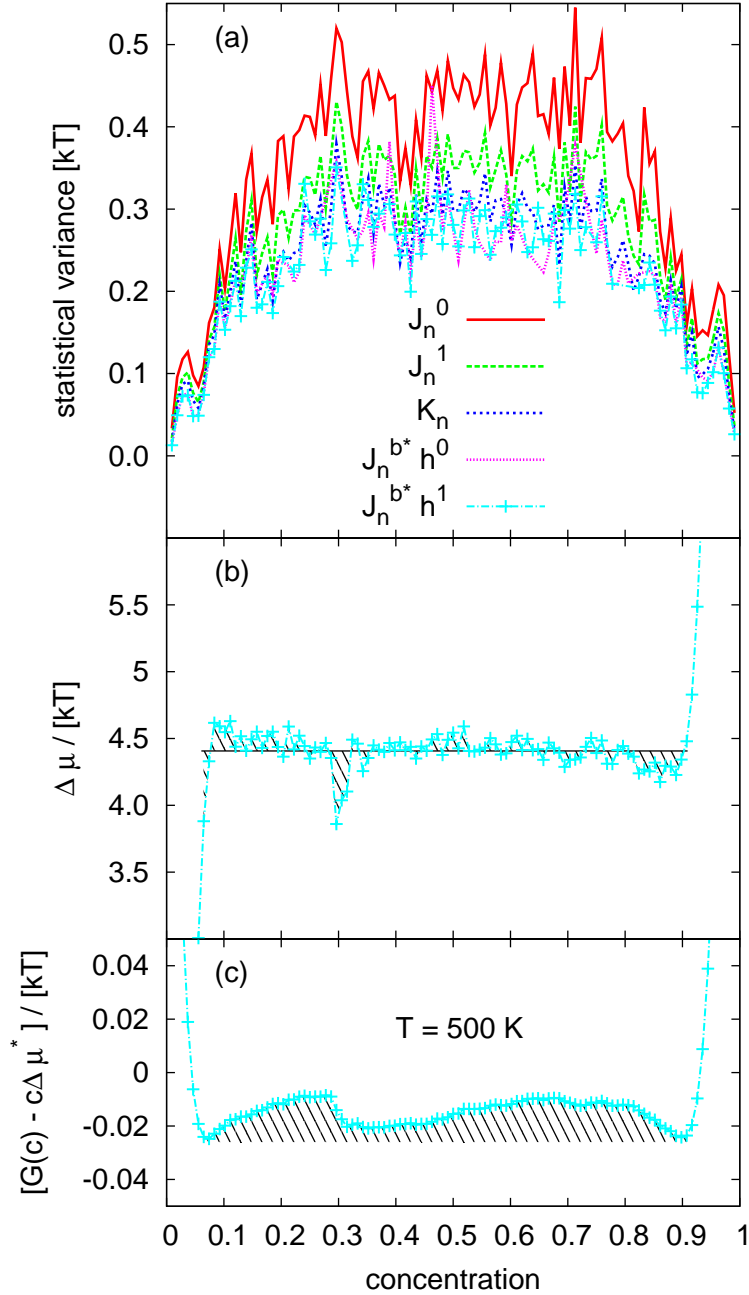}
  \caption{(a) statistical variances of estimated differences of
    excess chemical potentials [displayed in (b)];(c) solubility
    limits displayed in the concentration-temperature plane as
    obtained from Maxwell equal-area construction at $T=500$~K. 
    Used estimators are indicated in the legend.}
\label{fig:fecr}
\end{figure}

\Fref{fig:fecr} displays the statistical variances obtained for
the various estimators obtained with $n=400$ transmutations (per estimate) and
$m=200$ estimates at $T$=500~K ($T=(k \beta)^{-1}$). The aforementioned hierarchy still holds in
the present case for all concentrations. The Delmas-Jourdain estimator
still achieves better efficiency than the residence-weight estimator.
Because the amount of transmutations per estimate is much smaller than previously, it is more relevant to
estimate $b^\star$ using waste-recycling ($\mathrm{h}^1$) than without
($\mathrm{h}^0$). The former variant further decreases the variance
associated with the estimation of $b^\star$ by 20 \% on average over
the latter one. Averaging over all temperatures, estimator $\mathcal{J}^{b^\star}_n$-$\mathrm{h}^1$ 
decreases the variance by a factor of 2 to 5/2 compared to $\mathcal{J}^{0}_n$ and by 20 to 30 \%
compared to $\mathcal{J}^{b^\star}_n$ with $\mathrm{h}^0$. 
The $\Delta \mu$-values displayed in~\fref{fig:fecr}b correspond to a single estimate 
obtained from the $8 \cdot 10^4=m \times n$ transmutations used to evaluate 
the statistical variance. 
Figures~\ref{fig:fecr}b and~\ref{fig:fecr}c illustrate the equal-area or common-tangent
constructions of Maxwell and corroborate the occurrence of phase coexistence. 

\begin{figure}
\includegraphics{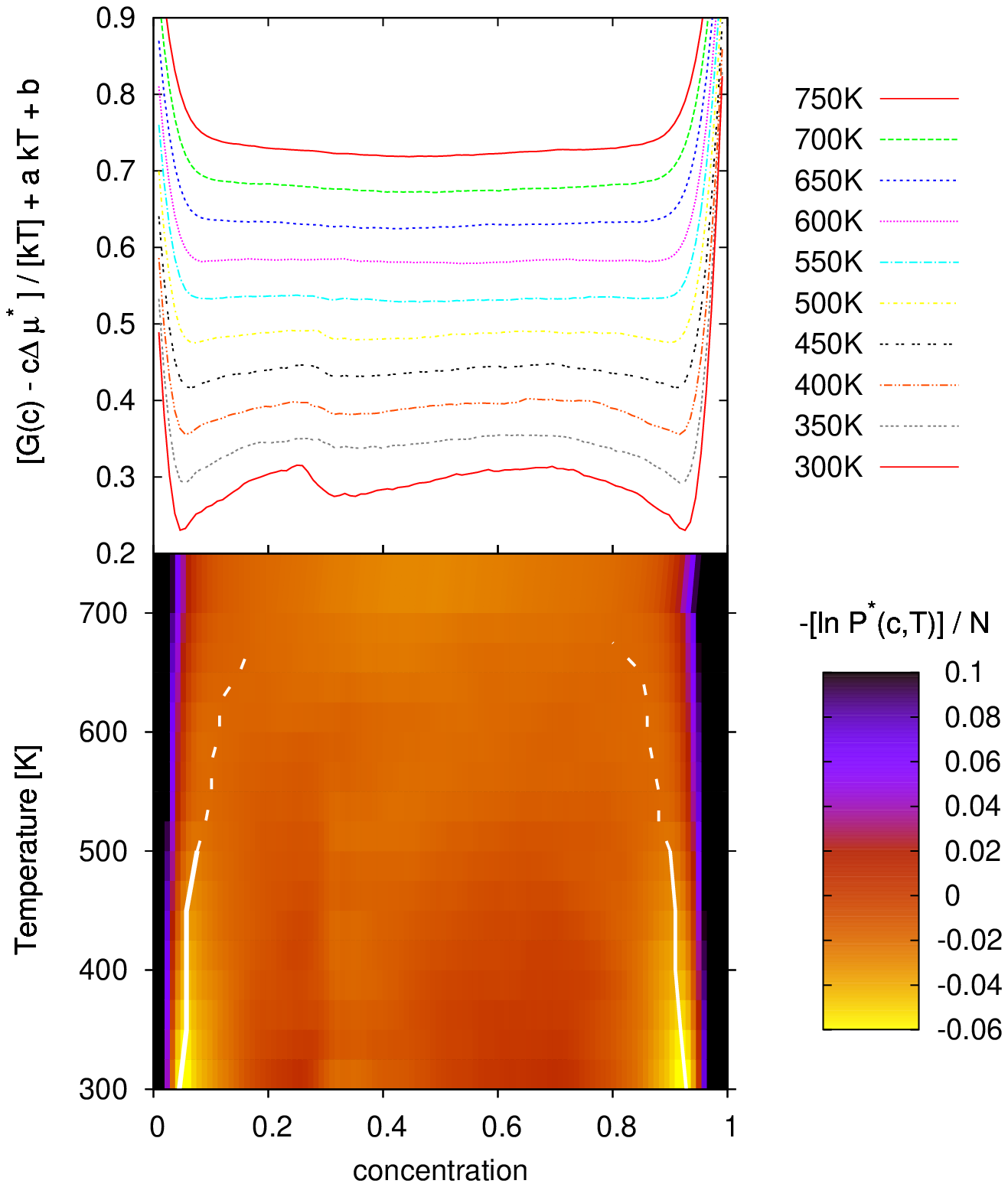}
  \caption{Top are the Gibbs free energies transformed for better
    visualization ($a=10^{-3}$ and $b=\ln Z /N $) as a function of
    concentration for various temperatures; bright regions
    on the bottom contour plot correspond to stability fields, the Fe
    and Cr solubility limits of the Fe-rich and Cr-rich phases are
    schematized by the two white lines. }
\label{fig:fecr_landscape}
\end{figure}

The reconstructed Gibbs free energy surface $G$ is less fluctuating because it is based on a
self-averaging integral \eref{eqn:gibbs_free_energy}. The potential
quantity $\Delta \mu^\star (T) $ in~\fref{fig:fecr}c is such that the occurrence
probabilities of the Fe-rich and Cr-rich phases are equal. The
occurrence probability is $\mathrm{P^\star}(c,T)=
Z^{-1}\re^{N\beta [c\Delta \mu^\star- G(c)]} $ where $Z = \int_0^1
\re^{N\beta [c^\prime\Delta \mu^\star-G(c^\prime)]} dc^\prime $ is the
semigrand canonical partition function~\cite{frenkel:2002} associated
with the finite computational cell. The potential function $\Delta \mu^\star (T)$ exactly coincides 
with the slope of the common-tangent at coexistence and in the thermodynamic limit only. 
Within finite computational cells, $\Delta \mu^\star (T)$ is defined even when there is no common tangent, 
is always easy to determine and fluctuates less than the slope of the common-tangent when there exists one. 
The transformed Gibbs free energy surfaces $G(c)-c \Delta \mu^\star$ are displayed in~\fref{fig:fecr_landscape}. 
The contour plot in the temperature-concentration plane of the bottom panel clearly visualizes
the miscibility gap in the composition range going approximately from 
Fe$_{0.1}$Cr$_{0.9}$ to Fe$_{0.9}$Cr$_{0.1}$. 

The coexistence lines (solubility limits) associated with the gap end around 500~K both in the iron-rich or
chromium-rich sides and are indicated by the solid white curve in~\fref{fig:fecr_landscape}b. 
Above 525~K, we observe that free energy profiles becomes lower for intermediate composition which indicates the presence of 
a stability field for these intermediate compositions and of two immiscibility fields for the Fe-rich and Cr-rich compositions. 
Solubility limits on the Fe-rich and Cr-rich sides associated with the two immiscibility fields are indicated by the dashed white curves in~\fref{fig:fecr_landscape}b. 
However, examination of snapshots of the simulated systems within the expected stability field still shows unmixed microstructures with a tendency to phase separation decreasing with increasing temperature from 525~K to 1700~K. This is attributed to the fact that the correlation length diverges at the critical temperature $T_c$ where the thermodynamic transition is second order. Given that the computational cell contains only $2\times 6^3$ atoms and that atomic interactions up to 7th nearest neighbors were shown to play an important role,~\cite{pareige:2009} strong finite-size effects are expected. The low temperature of 500~K measured for the present closure of the miscibility gap should therefore not be interpreted as an estimate of the critical temperature $T_c$ whose experimental value is expected to lie around 900~K.~\cite{bonny:2009} A finite-size scaling analysis~\cite{landau:2000} should therefore be carried out, which entails performing simulations with much larger computational cells. To achieve this task, the overall computation time has to be reduced considerably, possibly by evaluating the interatomic potential and forces on parallel computer architectures.~\cite{kim:2011} 
While the preliminary results presented here show that a direct and accurate construction of the equilibrium phase diagram of FeCr alloy is achievable in principle, they also emphasize the need for more extensive free energy calculations in this system. 

\section{Concluding remarks}

In this study, the optimal estimator proposed by Delmas and Jourdain
for waste-recycling Monte Carlo has been assessed numerically. As a
testbed, we implemented the mono-proposal Barker and
Metropolis-Hasting algorithms in transmutation ensembles of two alloy
systems simulated via the weighted work path ensemble.
We simultaneously estimated the free energy differences from the
nonequilibrium works measured along the transmutations. We find that
the estimator indeed achieves variance reduction compared to the other
Monte Carlo estimators that are compatible with the present sampling
approach. Furthermore, the maximal reduction of the statistical
variance that is predicted by the theory is attained for relatively
short simulations ($2 \cdot 10^3$ sampled paths).

Two other important methodologies have been proposed in the literature
for optimally extracting free energy differences from nonequilibrium
work measurements. Because they both require implementing specific
sampling schemes not suited to the present alloy systems, we did not
make any numerical comparisons. The differences in the implementation
can nevertheless be emphasized.

The first methodology, proposed by Crooks~\cite{crooks:2000} and
extended in Refs.~\onlinecite{maragakis:2008}
and~\onlinecite{minh:2008}, entails sampling the reference and target
thermodynamic states independently while performing bidirectional
nonequilibrium work measurements starting from the sampled states. One
then resorts to the optimal estimator of Bennett's acceptance ratio
method.~\cite{bennett:1976} A post-processing procedure constructs the
optimal estimator from Bayesian inference, neglecting correlations
between data.~\cite{chodera:2008} As the procedure minimizes the
variances obtained from a finite sample with respect to some control
parameter, the resulting estimator exhibits a statistical bias which
becomes negligible only in the limit of large samples, 
whereas the optimal estimator of Delmas and Jourdain may be stably
constructed when not in this limit and is always unbiased. 

One limitation of this alternative method relates to the lack of diagnostic tool to check 
the overlap between the forward and reverse switching processes. 
Insufficient overlap may result in poorly converged estimates. 
Another limitation lies in excluding the possibility of swapping
between the reference and target states.  For instance, when the
target state corresponds to a low temperature structure that traps the
system in a few basins of attraction, a single simulation of the
target system may not fully sample all of these basins.~\cite{adjanor:2006} 
Swapping between the reference and target state as employed in this paper can
avoid such non-ergodic sampling.

The second methodology~\cite{oberhofer:2008,LRS2010} entails
constructing an optimal biasing distribution rather than optimally
post-processing the data obtained in various thermodynamic states. The
optimal biasing distribution is derived analytically by minimizing an
asymptotic statistical variance given by the central limit
theorem.~\cite{chen:1997} 
  This approach possesses a number of limitations which we argue that waste recycling on work-weighted
  transmutation ensembles does not. First of all, the central limit theorem assumes independent and identically distributed random variables while sampled data are inevitably correlated; the denominator of the
  Delmas-Jourdain control parameter $b$ accounts for these
  correlations. Secondly, the rejected trial paths are discarded from
  the optimization procedure,
  potentially losing valuable information. Thirdly, the
  shape of the optimal distribution crucially depends on the free
  energy difference that one wishes to compute and cannot be
  analytically estimated \emph{a priori}. And finally, the optimal
  bias distribution constructed in this methodology is bimodal with
  respect to the work with its two
  peaks corresponding to the forward and reverse work distributions
  and with a density depletion in the overlapping region of the two
  latter distributions. This may result in highly correlated data and
  poor statistics compared to our work-weighted transmutation ensemble
  simulations that sample a unimodal
  distribution exhibiting good overlapping properties in term of work
  with respect to the distributions obtained with both forward and
  reverse switching. 
  Overall, we find that the Delmas-Jourdain estimator is unbiased and
  optimal in terms of asymptotic variance with respect to a simple
  control variate $b^\star$ that can be accurately estimated from the
  correlations in the collected data. One substantial benefit to
  employing this optimal estimator is the flexibility in the choice of
  the sampler, allowing us to couple waste recycling Monte Carlo with
  a work-weighted transmutation path ensemble ideally formulated to
  study the phase coexistence of binary alloys. However, in other
  scenarios, another Monte Carlo sampling scheme might be better
  posed, yet the optimal estimator of Delmas and Jourdain would still
  be applied quite similarly.
Since information associated with trial switching processes is optimally included in the free-energy estimates for any
sampler, this one is to be chosen so as to facilitate the phase space exploration. 

Nonetheless, our approach employing the optimal estimator of Delmas
and Jourdain shares a qualitative feature with the methodology based
on the optimal biasing distribution.~\cite{oberhofer:2008} 
Indeed, forward and reverse trial paths, which contribute substantially to the free energy estimate given by the Delmas-Jourdain estimator for the measured optimal values of the control variate $b^\star$, concommitantly cover the two peaks of the optimal bias distribution.

Concerning the investigated examples of binary alloys, we point out
that achieving numerical ergodicity in the reference and target
thermodynamic states entails exchanging the allocation of atoms of
distinct types on the underlying lattice (which amounts to performing pairs 
of simultaneous transmutations with opposite direction) with a high
enough frequency. With path-sampling, exchanges of atom allocation are
automatically achieved when trial paths are successively accepted
within the transmutation ensemble (as illustrated in
\fref{fig:schematic}). It turns out that the phase space exploration
is considerably facilitated. Resorting to such a path-sampling scheme
was found particularly advantageous in the FeCr alloy system which
presents a large magnetic misfit. Direct exchange
moves sampled using a standard scheme in the reference and target
system would have been extremely infrequent and thus computationally
expensive compared to the gradual transmutations considered here. 
The approach might work as well in the numerous alloy systems 
exhibiting large atomic misfits. 

Further assessments of the estimator should consider more general
scheduling for the nonequilibrium dynamics rather than simply
\eref{eqn:evolution} and a multi-proposal sampling scheme. Optimal
waste-recycling estimators can possibly be derived with multiple
proposals based on the same control variate problem
structure.~\cite{delmas:2009} Work in this direction is in
progress.~\cite{kim:2011}

\begin{acknowledgments} 
  We warmly thank Benjamin Jourdain for
  suggesting that we test the optimal estimator and for advising us.
  Stimulating discussions with Berend Smit, Christoph Dellago and
  Peter Bolhuis are also gratefully acknowledged.
\end{acknowledgments}

\appendix

\section{Residence-weight estimator~\label{app:inference}}

In the context of path-sampling, waste-recycling estimators may also
be derived from statistical inference. The algorithm indeed samples a
marginal probability distribution associated with the Bayesian
prior $\rP_\theta$. As a result, information relative to unselected trajectories or to states contained in the
trajectories themselves may be inferred online from a Bayesian posterior likelihood without post-processing.
The marginal probability density of the sampled web $\left\{ z ,
  \tilde{z} \right\}$ is \begin{equation}
 \Phi_\theta \left( \left\{ z , \tilde{z} \right\} \right) = \frac{1}{2} \left[ q( \left\{ z, \tilde{z} \right\} | z) \rP_\theta (z) + q( \left\{ z , \tilde{z} \right\}  | \tilde{z}) \rP_\theta (\tilde{z}) \right] .  
\end{equation}
recalling that $q$ is the conditional probability to construct
$\left\{ z , \tilde{z} \right\}$ from either $z$ or $\tilde{z}$. The
factor $1/2$ acts as a normalizing constant and relates to the
position of the prior $z$ in $\left\{ z , \tilde{z} \right\}$. $\rP_\theta(z)$ and $\rP_\theta(\tilde{z})$ are the prior
probabilities of $\left\{ z , \tilde{z} \right\}$.

With the Boltzmann sampler, the rejection probability of $\tilde{z}$
is $1-p^B(\tilde{z}|\left\{z,\tilde{z} \right\})$. It is equal to the
posterior likelihood $p^B(z | \left\{ z , \tilde{z} \right\} )$ of $z$
and to the acceptance probability of $z$ constructed from $\tilde{z}$
using the proposal probability $q_\alpha(\left\{ z , \tilde{z}
\right\} | \tilde{z})$. Posterior likelihoods relate to marginal and
conditional probabilities via Bayes theorem \begin{equation}
 p^B(z_c|\left\{ z , \tilde{z} \right\} )= \frac{ \frac{1}{2} q(\left\{ z , \tilde{z} \right\} | z_c) \rP(z_c)}{\Phi_\theta(\left\{ z , \tilde{z} \right\} )}
\end{equation}
where $z_c$ is either $z$ or $\tilde{z}$. As a result, the Boltzmann
sampler leaves invariant the probability distribution $\Phi_\theta
\left(\{ z , \tilde{z} \}\right)$, since it satisfies the detailed
balance equation ($z_c \in \left\{ z , \tilde{z} \right\} \cap \left\{
  z^\prime , \tilde{z}^\prime \right\} $)
\begin{equation}
q(\left\{ z^\prime , \tilde{z}^\prime \right\} | z_c) p^B( z_c| \left\{ z , \tilde{z} \right\}) \Phi_\theta( \left\{ z , \tilde{z} \right\}) = q( \left\{ z , \tilde{z} \right\} | z_c) p^B( z_c | \left\{ z^\prime , \tilde{z}^\prime \right\}) \Phi_\theta(\left\{ z^\prime , \tilde{z}^\prime \right\}).~\label{eqn:BDgeneral}
\end{equation}
In \eref{eqn:BDgeneral}, both $p^B(z_c|\{z^\prime,\tilde{z}^\prime \})$ and
$p^B(z_c|\{z^\prime,\tilde{z}^\prime \})$ correspond to either an acceptance or a rejection probability.

Let ${\Gamma_\alpha}$ denote the space of webs sharing a common state
at $t=\alpha \tau$. The total web space is $\Gamma_0 \cup \Gamma_1$
and we have equipartition of webs in $\Gamma_0$ and $\Gamma_1$. The
free energy differences can be cast in the form~\cite{athenes:2007}
($\alpha \in \{0 ,1\}$) \begin{eqnarray}
\re^{g_\theta - g_\alpha} & = & \int_{\Gamma_\alpha} q_\alpha(\{ z ,\tilde{z} \}| z) \rP_\theta(z) \re^{\beta(\theta-\alpha)W(z)} + q_\alpha(\{ z ,\tilde{z} \} | \tilde{z}) \rP_\theta(\tilde{z}) \re^{\beta(\theta-\alpha)W(\tilde{z})} \cD z \cD \tilde{z} \nonumber \\
 & = & \int_{\Gamma_\alpha} \left[ p^B_\alpha( z| \{ z ,\tilde{z} \}) \re^{\beta(\theta-\alpha)W(z)} + p^B_\alpha \left(\tilde{z} | \{ z ,\tilde{z} \} \right) \re^{\beta(\theta-\alpha)W(\tilde{z})} \right]\Phi_\theta(\{ z ,\tilde{z} \}) \cD z \cD \tilde{z} \nonumber \\
 & = & 2 \int_{\Gamma_\alpha} \left[ \re^{\beta(\alpha-\theta)W(z)} + \re^{\beta(\alpha-\theta)W(\tilde{z})} \right]^{-1}\Phi_\theta(\{ z ,\tilde{z} \}) \cD z \cD \tilde{z} \nonumber
\end{eqnarray}
Since the algorithm sample the distribution $\Phi_\theta$, the free energy difference $g_\theta - g_\alpha$ can be estimated from the logarithm of the following estimator 
\begin{eqnarray}
K_n\left( \re^{\beta(\theta - \alpha)W} \right) & = & \frac{4}{n} \sum_{k=1}^{n/2} 
\left[ \re^{\beta(\alpha-\theta)W(z_{2k+\alpha})}+\re^{\beta(\alpha-\theta)W(\tilde{z}_{2k+\alpha})} \right]^{-1}. \label{eqn:resi}
\end{eqnarray}
In \eref{eqn:resi}, trial paths $\tilde{z}_{2k}$ and $\tilde{z}_{2k+1}$
are generated forward and backward, respectively. Hence, we have
$(z_{2k+\alpha},\tilde{z}_{2k+\alpha}) \in \Gamma_\alpha$. For more
details, see Ref.~\onlinecite{athenes:2007}.

\bibliography{apssamp}

\begin{thebibliography}{00} 

\bibitem{frenkel:2006} D. Frenkel, Waste-recycling Monte Carlo, in ``Computer Simulations in Condensed Matter Systems'', Lect. Notes Phys. {\bf 703}, 127 (2006).  

\bibitem{ceperley:1977} D. Ceperley, G. Ghester and M. Kalos, {Phys. Rev. B} {\bf 16}, 3081 (1977). 

\bibitem{frenkel:2004} D. Frenkel, Proc. Natl. Acad. Sci. U.S.A. {\bf 101}, 17571 (2004).   

\bibitem{coluzza:2005} I. Coluzza and D. Frenkel, {ChemPhysChem} {\bf 6 } 1779 (2005). 

\bibitem{athenes:2008c} M. Ath\`enes and F. Calvo, ChemPhysChem {\bf 9}, 2332 (2008). 

\bibitem{athenes:2002b} M Ath\`enes, Phys. Rev. E, {\bf 66}, 046705 (2002). 

\bibitem{athenes:2004} M. Ath\`enes, {Eur. Phys. J. B}, {\bf 38}, 651 (2004). 

\bibitem{athenes:2002a} M Ath\`enes, Phys. Rev. E {\bf 66}, 016701 (2002). 

\bibitem{boulougouris:2005} G. Boulougouris and D. Frenkel, J. Chem. Theory Comp. {\bf 1}, 389 (2005). 

\bibitem{athenes:2010} M. Ath\`enes, M.-C. Marinica, J. Comput. Phys. {\bf 229}, 7129 (2010).  

\bibitem{athenes:2007} M. Ath{\`e}nes, {Eur. Phys. J. B}, {\bf 58}, 83, (2007). 
 
\bibitem{delmas:2006} J.-F. Delmas, B. Jourdain, arXiv:math/0611949v1. 

\bibitem{delmas:2009} J.-F. Delmas, B. Jourdain, J. Appl. Probab. {\bf 46}, 938 (2009). 

\bibitem{frenkel:2002} D. Frenkel and B. Smit, Understanding molecular simulation, Academic Press, New York 2002, p389. 

\bibitem{LRS2010} T. Leli{\`e}vre, M. Rousset and G. Stoltz, Free-energy computations: a mathematical perspective, Imperial College Press, 2010. 
 
\bibitem{Crooks:1998} G. Crooks, J. Stat. Phys. {\bf 90}, 1481 (1998). 

\bibitem{crooks:2000} G. Crooks, Phys. Rev. E {\bf 61}, 2361 (2000). 

\bibitem{hummer:2001} G. Hummer and A Szabo, Proc. Natl. Acad. Sci. U.S.A. {\bf 98}, 3658 (2001). 

\bibitem{laio:2002} A. Laio, M. Parrinello, Proc. Nat. Acad. Sci. USA {\bf 99}, 12562 (2002). 

\bibitem{jarzynski:1997} C. Jarzynski, Phys. Rev. Lett. {\bf 78}, 2690 (1997).  

\bibitem{jarzynski:2008} C. Jarzynski, Eur. Phys. J. B {\bf 59}, 331 (2008). 

\bibitem{adjanor:2005} G. Adjanor  and M. Ath\`enes, J. Chem. Phys. {\bf 123}, 234104 (2005). 

\bibitem{sun:2003} S. X. Sun, J. Chem. Phys.  {\bf 118}, 5769 (2003). 

\bibitem{oberhofer:2005} H. Oberhofer, C. Dellago and P. L. Geissler, J. Phys. Chem. B {\bf 109}, 6902 (2005). 

\bibitem{lechner:2007} W. Lechner and C. Dellago, J. Stat. Mech. {\bf 04}, P04001 (2007). 

\bibitem{jarzynski:2007} C. Jarzynski, C. R. Physics {\bf 8}, 495 (2007). 

\bibitem{horowitz:2008} J. Horowitz and C. Jarzynski, {J. Stat. Mech.: Theory Exp.} {\bf 11}, P11002 (2007). 

\bibitem{bochkov77} G N Bochkov  and Yu. E. Kuzovlev, J. Exp. Theor. Phys. {\bf 72}, 238 (1977). 

\bibitem{bochkov81} G N Bochkov and Yu. E Kuzovlev, Physica A {\bf 106}, 443 (1981). 

\bibitem{ytreberg:2006} F.M. Ytreberg, R.H. Swendsen, D.M. Zuckermann, J.Chem. Phys. {\bf 125}, 184114 (2006). 

\bibitem{duflo:1997} M. Duflo. Random iterative models, volume 34 of Applications of Mathematics (New York). Springer-Verlag, Berlin, 1997. 

\bibitem{ducastelle:1991} F. Ducastelle, Order and Phase Stability in Alloys, Elsevier Science Publisher, North Holland, 1991. 

\bibitem{athenes:2000} M. Ath\`enes, P. Bellon and G. Martin, Acta Mat. {\bf 48}, (2000) 2675. 

\bibitem{olsson:2005} P. Olsson, J. Wallenius, C. Domain, K. Nordlund, L. Malerba, Phys. Rev. B {\bf 72}, 214119 (2005) and  Phys. Rev. B {\bf 74}, 229906 (2006). 

\bibitem{bonny:2009} G. Bonny, R. Pasianot, L. Malerba, A. Caro, P. Olsson, M. Yu. Lavrentiev, J. Nucl. Mater. {\bf 385}, 268 (2009). 

\bibitem{pareige:2009} C. Pareige, C. Domain and P. Olsson, J. Appl. Phys. {\bf 106}, 104906 (2009). 

\bibitem{landau:2000} D. Landau and K. Binder, A guide to Monte Carlo simutations in Statistical Physics, Cambridge University Press (2010). 

\bibitem{kim:2011} J. Kim, J. M. Rodgers, M. Ath\`enes, and B. Smit, in preparation. 

\bibitem{maragakis:2008}  P. Maragakis, F. Ritort, C. Bustamante, M. Karplus, and G. Crooks, J. Comp. Phys. {\bf 129}, 024102 (2008). 
%
\bibitem{minh:2008} D. Minh and A. Adib, Phys. Rev. Lett. {\bf 100 }, 180602 (2008). 

\bibitem{bennett:1976} C.~H. Bennett, J. Comp. Phys. {\bf 22 }, 245 (1976). 

\bibitem{chodera:2008} M. Shirts and J. Chodera, J. Chem. Phys. {\bf 129}, 124105 (2008). 

\bibitem{adjanor:2006} G. Adjanor, M. Ath\`enes, and F. Calvo, Eur. Phys. J. B {\bf 53}, 47 (2006). 

\bibitem{oberhofer:2008} H. Oberhofer and C. Dellago, Comp. Phys. Com. {\bf 179}, 41 (2008). 

\bibitem{chen:1997} M.-H. Chen and Q.-M. Shao, Ann. Stat. {\bf 25}, (1997) 1563. 

\end{thebibliography}

\end{document}